# Unlock Anionic Behavior of Calcium Through Pressure Engineering

Yang Lv,[1] Junwei Li,[2] Jianfu Li,[1,*] Yong Liu,[1] Jianan Yuan,[1] Jiani Lin,[1] Saori Kawaguchi-Imada[3], Qingyang Hu,[2,*] and Xiaoli Wang[1,*]

[1]*School of Physics and Electronic Information, Yantai University, Yantai 264005, China*

[2]*Center for High Pressure Science and Technology Advanced Research, Beijing 100193, China*

[3]*Japan Synchrotron Radiation Research Institute, Hyogo, Japan*

Corresponding authors: jianfuli@ytu.edu.cn (Jianfu Li); qingyang.hu@hpstar.ac.cn (Qingyang Hu); xlwang@ytu.edu.cn (Xiaoli wang)

## Abstract

An isolated calcium (Ca) atom has empty *d*-orbitals under ambient conditions. However, *s*-*d* band hybridization has been observed in both elemental Ca and compounds by manipulating thermodynamic conditions. Here, we reveal that the Ca 3*d*-band can even capture electrons from halogen atoms under pressure, exhibiting anionic behaviors in iodides. We predict a CsCl-type monovalent CaI at above 50 GPa by employing first-principles structural searching and successfully identified the phase at 84 GPa using *in situ* X-ray diffraction. We further reveal that, due to the effect of orbital broadening, unusual charge transfer from the 5*p* orbitals of I to the 3*d* orbitals of Ca in CaI, gradually reverses the ionicity of Ca and becomes the anionic ICa at 485 GPa. Multivalent Ca stabilizes a set of metallic iodides with eight- to ten-fold iodine hyper-coordination. Our findings demonstrate that the valence states of Ca can vary from negative to +2, suggesting much greater complexity of Ca chemistry under ultrahigh pressures.

## Introduction

The stability, bonding, and chemical reactivity of elements are primarily determined by the outermost shell of electrons, known as the valence electrons[1–3]. Under ambient conditions, the valence electrons of elements are organized and even predicted on the periodic table, which illustrates how electrons fill the atomic shells. However, this fundamental chemistry understanding requires adjustment under gigapascal (GPa) pressure regime, since pressure can selectively shift the energy of atomic orbitals[4].

An outstanding example of this phenomenon is the pressure-induced $s \rightarrow d$ electron transition. Originally discovered in alkali and alkaline-earth metals[5–7], this transition was later observed in transition and rare Earth metals such as Sc, La, and even noble gases[8]. In principle, $s \rightarrow d$ electron transition occurs due to the energy collapse of *d* orbitals under high-pressure conditions[9], where electrons residing in the outer *s*-block (and sometimes *p*-block) shell are squeezed into the lower-energy, inner *d* orbitals. It is historically used by crystallographers to explain the symmetry breaking observed in alkali or alkaline-earth metals under pressure, despite being later identified as a minor effect compared to electron localization[10]. Regardless of its role in crystal symmetry, the lowered energies of *d* orbitals cause drastic changes in elemental characteristics. For example, in pressurized metals like K, Ca, and Ga, filling 3*d* orbitals would be more energetically favorable with increasing pressure. This pressure-engineered effect may deviate from the strict hydrogenic ordering typically prescribed by the Aufbau principle[11]. At 31 GPa, K with occupied 3*d* electrons exhibits a smaller atomic radius, and forms K-Ni compounds after heating to 2500 K[12]. Similarly, the reduced energy 3*d* orbitals in confined Ca can substantially decrease charge transfer within Ca-Te compounds, causing Ca to lose its metallicity[13].

We focus on the chemistry of Ca in this work, as Ca is the last element with a vacant *d*-orbital. Due to the Coulomb repulsion exerted by core electrons on valence electrons, single-element calcium counterintuitively transits from a metal to a semiconductor under pressure[14,15]. Within the pressure range of 0–32 GPa, its coordination number and packing efficiency decrease, dropping from 12 to 6 and from 0.74 to 0.52, respectively[16,17]. High-pressure Ca allotropes also include incommensurate Ca-IV and host-guest structured Ca-V[18,19], and eventually become a superconductor at 161 GPa with a transition temperature as high as 25 K[20]. In compounds, the traditional understanding that calcium ions can only exist in the +2 valence state has been challenged by the discovery of monovalent calcium ions, and anionic Ca in organic hydride complexes[21]. For instance, CaCl crystals containing monovalent calcium ions were characterized in reduced graphene oxide, exhibiting intriguing properties such as metallicity, ferromagnetism, and piezoelectric-like effects[22]. This monovalent calcium chloride was also synthesized under high-pressure conditions of 50 GPa[23]. More non-stoichiometric super-chlorides were predicted by theory, such as $Ca_2Cl_{11}$ which features partial monovalent calcium anions[24]. These examples indicate that calcium can manifest more intriguing and remarkable properties within compounds, but anionic Ca in inorganic salt has not been reported yet.

In this work, we combine high-pressure-temperature synthesis, *in-situ* x-ray diffraction, and first-principles simulation to investigate the evolution of Ca valence state within the Ca-I binary system. We propose Ca can even be stabilized in the negative valence state as iodine "calcium-ite" (ICa), thus acting as a nonmetal oxidizing agent. Given iodine's larger electronegativity difference with Ca[25], we presume they form $Ca_xI_y$ ionic compounds. However, pressure-induced reduction of $Ca^{2+}$ into $Ca^{1+}$ and even anionic Ca were observed through pressure engineering. The variety of Ca valence states open up opportunities for the discovery of non-stoichiometric Ca iodides.

## Results

### Synthesizing monovalent CaI under pressure.

The chemical stabilities of $Ca_xI_y$ compounds with $y/x$=1/3, 1/2, 1, 2, 3, and 4 are evaluated by calculating their formation enthalpies ($\Delta H$) at static conditions up to 800 GPa. The energies are then related to the products of their dissociation into constituent elements, as summarized in Figure 1(a). The compounds located on the convex hull are stable against decomposition into other compositions. In contrast, compounds above the convex hull are unstable or metastable if the kinetic barrier is sufficiently high. At ambient pressure, all calculated Ca-I compounds except $CaI_2$ have negative $\Delta H$, following the conventional ionic rule. In the extended 50 ~ 800 GPa pressure range, non-stoichiometric iodides such as CaI, $Ca_2I$, and $Ca_3I$ emerged on the convex hull (Figure 1b and Figure 3). Their lattice parameters are given in Table S1, and the calculated phonon dispersion curves suggest that these structures are dynamically stable (Figure S6- S9). Notably, the CaI (*Pm*-3*m*) phase has consistently exhibited the lowest formation enthalpy among all structures starting from 50 GPa.

We then conducted high-pressure experiments to synthesize CaI with laser heating. The experiment started from the stoichiometric $CaI_2$, which is commercially available (CAS: 10102-68-8). Preparation steps of samples in diamond anvil cells are found in the Methods section. The sample was compressed up to 84 GPa, and was heated at two pressure-temperature (*P*-*T*) pairs of 44 GPa-1600 K and 84 GPa-2200 K at the beamline 10XU of SPring-8, Japan. X-ray diffraction patterns were taken after quenching to ambient temperature (Figures 1c and 1d). At 44 GPa, $CaI_2$ maintained its stoichiometry but transited to the high-pressure phase with space group *P*-62*m*, consistent with theory (Figure 1a). Its lattice constants are refined as $a$ = 7.762(20) Å, $c$ = 3.683(22) Å, compared with first-principles simulation of $a$ = 7.742 Å and $c$ = 3.594 Å. At 84 GPa, we observed a new set of diffraction peaks associated with the monovalent CaI (CsCl-type, B2) to emerge and coexist with $CaI_2$. The lattice constant of CaI at 84 GPa is $a$ = 3.360(7) Å, slightly greater than the predicted value of 3.241 Å. This observation of CaI marks the decomposition of $CaI_2$ into CaI and I, reducing the valence state of $Ca^{2+}$ to $Ca^{1+}$.

**Charge transfer and anionic behavior of Ca**

The prediction and synthesis of CaI set up the foundation for analyzing charge transfer between Ca and I ions. For this archetypal B2 structure, atoms occupy either the center or corners of a cubic lattice. We therefore performed a series of computational analyses on the crystal and electronic structures of B2-CaI and representative $CaI_2$ phases identified from the convex hull diagram. Specifically, we calculated the projected density of states (PDOS), the net charge between Ca and I atoms using the Bader charge[26–28], the difference of charge density, and band occupation. As Figure 2(a) illustrates, the semiconducting $CaI_2$ metallizes during the decomposition to CaI at 50 GPa. The Ca-3*d* and -4*s* orbitals of $CaI_2$ remain unoccupied, resulting in $Ca^{2+}$ as a reducing agent. However, the Ca-3*d* orbitals of CaI is activated and their orbital contributions near and below the Fermi level become more pronounced, indicating lowered orbital energy and the tendency to capture electrons. This trend deepens with pressure, and since CaI has extended thermodynamic stability up to 800 GPa, we calculate the activation of the Ca-3*d* orbitals in multi-megabar regions. These extreme conditions were commonly found in the deep interiors of exoplanets such as super Earths. At above 485 GPa, the electron occupancy of the Ca-3*d* orbitals near the Fermi level surpasses that of the I-5*p*, suggesting that electrons originally populated in the I-5*p* orbitals migrate to the Ca-3*d* orbitals. Comparing the DOS of Ca-3*d* and I-5*p* orbitals under varying pressures, we reveal that the observed inter-orbital electron transfer is driven by pressure-induced orbital broadening. Specifically, as shown in Figure 4(a), such broadening extended the Ca-3*d* orbitals crossing the Fermi level and gained electrons. In contrast, the I-5*p* orbitals cross the Fermi level rightward and lose electrons.

From Figure 2(b), the charge transfers between Ca and I decrease progressively with increasing pressure. The net charges of Ca and I eventually cross at 485 GPa, indicating neutral ionicity. Beyond this critical pressure, Ca atoms exhibit an intriguing metal-nonmetal transition, start to accept electrons, and adopt negative states. Conversely, I atoms act as electron donors. Given that low-lying Ca-3*d* orbitals tend to hybridize with non-metallic ions and induce errors[29,30], we applied an external potential to Ca-3*d* orbitals (Figure S5) to validate our results[31,32]. The findings show that while the metal-nonmetal transitions pressure increased by 252 GPa, the anionic tendency of Ca remained unaffected. Notably, at 800 GPa, the Bader charge of Ca reaches -0.55. Considering the well-known underestimation of Bader charges, the actual valence state may be even lower. Figure 2(c) plots the charge density difference at 200, 400, and 800 GPa, in which electrons originally aggregated around I transfer to and accumulate around Ca at higher pressures. This is consistent with our calculated band occupation of B2-CaI as shown in Figure 2(d). With increasing pressure, Ca-3*d* orbitals become more energetically favored due to the augmented capacity of the low-energy inner-shell of the *d* orbitals to hold electrons[33]. This, in turn, gives rise to a disparity in the pressure sensitivity between these orbitals. In the meantime, the energy of the I-5*p* orbitals shows a notable increase as pressure rises, consistent with literature[4]. In summary, these calculated electronic structures unambiguously confirm the anionic behavior of nonmetal Ca in attracting electrons.

**Expanded Ca chemistry under pressure with Ca-I hyper-coordination**

The Ca chemistry with -1, 0, +1, and +2 multivalent states over an extended pressure range leads to novel Ca-I stoichiometries and phases. Besides the reported $CaI_2$ and CaI, our structural searching algorithm also found $Ca_2I$ and $CaI_3$ featuring complex Ca-I polyhedral structures with 8-10 coordination numbers (Figure 3). For example, in the most I-enriched $CaI_3$, each Ca atom is coordinated with 10 I atoms, forming $CaI_{10}$ bicapped square antiprisms. These hyper-coordinated structures and their corresponding thermostability fields are included in the supplementary materials. Their formations are possibly explained by the decomposition of iodides under high pressure, which was mentioned elsewhere[34–38].

We further establish the relation between the I-concentration and the net charge in Figure 2(e) to compare the anionicity of Ca in these non-stoichiometric compounds. These high-coordinated, I-enriched compounds are more anionic (lower charge) at pressures below 700 GPa. However, at higher pressures, nonmetal Ca in B2-CaI exhibits the lowest charge among all stoichiometries, possibly due to the higher atomic packing efficiency of B2-ICa than other I-enriched components. The smaller average atomic volume in the B2-ICa indicates it is more chemically compressed[39,40]. We therefore speculate that the crystal and electronic structures are highly relevant to of nonmetal Ca, which will be discussed in the next paragraph.

Our electronic calculation highlights the role of Ca-3*d* band broadening in developing Ca's anionic behavior (Figure 4(a)). By fixing the composition as CaI, we further analyze the relation between net charge and chemical pressure. Here, the chemical pressure is indicated by Ca-I bond length, which is controlled by the space group and the ionic bond distance and built upon various low-energy CaI structures found by our structural searching algorithm. This chemical pressure versus net charge plot is divided into two distinct regions (Figure 4(b)). Structures with equal Ca-I bond lengths establish almost linear relation with its net charge, while structures with extensive Ca-Ca and I-I coordination appear above the linear line. Such interaction between the same elements hinders Ca-I reverse charge transfer, causing upward deviation from linearity. While only B2-CaI remains stable on the convex hull under high pressure due to the contribution of the $\Delta PV$ term Figure S2, other shorter-bond-distance phases are thermodynamically unstable. Given these findings and the shared partially filled 3*d* orbitals of transition metals, we hypothesize that the B2 structure is more energetically favored in a variety of monovalent metal iodides under high-pressure. When occupied and empty orbitals lie near the Fermi level, pressure-induced broadening drives orbital hybridization, reducing ionicity and inducing metal-nonmetal transitions in 3*d* metals.

## Conclusion

In this work, to obtained the high-pressure phase diagram of Ca-I system and structural information in multy-stoichiometries, we employed first-principles calculations and an unbiased structure-searching technique, in combination with *in situ* X-ray diffraction and a laser-heating diamond anvil cell. We have discovered and confirmed that pressure activated the Ca-3*d* orbital by orbital broadening. This leads an unusual charge transfer from the I-5*p* orbitals to the Ca-3*d* orbitals in CaI, gradually reverses the ionicity of Ca and becomes the anionic ICa at 485 GPa. This reverse also leads Ca-I compounds' coordination numbers and structures, as the iodine content increases, of $Ca_2I$, CaI, $CaI_2$, and $CaI_3$ increase to 5, 8, 9, and 10, respectively. Analysis of the net charge and chemical pressure (represented by the Ca-I bond length) indicates that a prevalence of homonuclear interactions (i.e., excessive Ca-Ca and I-I bonds) inhibits the formation of anionic Ca. In structures dominated by heteroanionic Ca-I bonding, the Ca-I bond length exhibits a nearly linear relationship with the net charge on the anion Ca. Considering the significant contribution from the $\Delta PV$ term enthalpy term under high pressure and the partially filled 3*d* orbitals of the transition metal, we have made a scientific hypothesis regarding the prevalence of the B2 phase in monovalent metal iodides under high pressure, and the feasibility of unlocking the anionic behavior of 3*d* metals.

**Acknowledgment**

This work was supported by the National Natural Science Foundation of China (Grants No. 11974154, 42150101, and 12304278), the Taishan Scholars Special Funding for Construction Projects (Grants No. tstp20230622), the Natural Science Foundation of Shandong Province (Grants No. ZR2022MA004, ZR2023QA127, and ZR2024QA121), and the Special Foundation of Yantai for Leading Talents above Provincial Level.

**Materials and Methods**

To analyze the preferred structures of Ca-I at high pressures, we used the Particle Swarm Optimization (PSO) algorithm[42,43] as implemented in the Crystal structure Analysis by Particle Swarm Optimization (CALYPSO) code[44,45]. This method can find the most stable structures just by knowing the chemical composition, therefore, it is unbiased by any previously known structure information. CALYPSO has been validated with various known systems, ranging from elements and binary compounds[35,37,46]. Total energy calculations are performed in the framework of density functional theory within the generalized gradient approximation[47] as implemented in the VASP code[48,49]. The electron−ion interaction is described by pseudopotentials built within the scalar relativistic projector augmented wave approximation with $3p^6 4s^2$ and $5s^2 5p^5$ valence electrons for Ca and I atoms, respectively. A cutoff energy of 700 eV and appropriate Monkhorst−Pack k-meshes are used to ensure that total energy calculations are converged to less than 1 meV/atom. The thermodynamic stability of different $Ca_mI_n$ compounds to elemental Ca and I solids at each pressure is evaluated through the below equation:

$$H_f\,(Ca_mI_n) = [H(Ca_mI_n) - mH(Ca) - nH(I)] \,/\, (m + n)$$

in which $H$ is the enthalpy of the most stable structure of a specified composition at the given pressure. To determine the dynamic stability of predicted structures, the phonon calculations were performed using the finite displacement approach as implemented in the Phonopy code[50]. To analyze the interatomic interaction, the crystal orbital Hamilton populations (COHP)[51] was calculated using the LOBSTER program[52,53].

The starting material, calcium iodide powder (Alfa Aesar, 99.5%), was stored and handled in an argon-filled glove box. The starting material was loaded into symmetric diamond anvil cells inside the glove box, and the sample chambers were then sealed and compressed to ~10 GPa before being moved outside the glove box to avoid exposure to the air. Subsequently, samples were further compressed to target pressures and then heated using a double-sided laser heating system described elsewhere[54]. Temperatures were determined by fitting the four monochromatic thermal emission spectra to the Wien function[55]. Pressures were determined before and after heating using the calibrated ruby fluorescence line shift[56] or the first-order Raman band shift of diamond anvils[57] acquired on a micro-Raman spectrometer system. We estimate uncertainty of pressure was up to ±3 GPa. The unit cell is calculated by indexing the diffraction peaks of the Ca-I compound by a nonlinear fitting program[58], more details of experiment are included in the supplementary materials.

## References

(1) Tsuneda, T. Chemical Reaction Analyses Based on Orbitals and Orbital Energies. *Int J of Quantum Chemistry* **2015**, *115* (5), 270–282. https://doi.org/10.1002/qua.24805.

(2) Mulliken, R. S. Bonding Power of Electrons and Theory of Valence. *Chem. Rev.* **1931**, *9* (3), 347–388. https://doi.org/10.1021/cr60034a001.

(3) Pauling, L. The Nature of the Chemical Bond-1992.

(4) Miao, M.-S.; Hoffmann, R. High Pressure Electrides: A Predictive Chemical and Physical Theory. *Acc. Chem. Res.* **2014**, *47* (4), 1311–1317. https://doi.org/10.1021/ar4002922.

(5) Xie, Y.; Tse, J. S.; Cui, T.; Oganov, A. R.; He, Z.; Ma, Y.; Zou, G. Electronic and Phonon Instabilities in Face-Centered-Cubic Alkali Metals under Pressure Studied Using *Ab Initio* Calculations. *Phys. Rev. B* **2007**, *75* (6), 064102. https://doi.org/10.1103/PhysRevB.75.064102.

(6) Maksimov, E. G.; Magnitskaya, M. V.; Fortov, V. E. Non-Simple Behavior of Simple Metals at High Pressure. *Phys.-Usp.* **2005**, *48* (8), 761–780. https://doi.org/10.1070/PU2005v048n08ABEH002315.

(7) Ahuja, R.; Eriksson, O.; Wills, J. M.; Johansson, B. Theoretical Confirmation of the High Pressure Simple Cubic Phase in Calcium. *Phys. Rev. Lett.* **1995**, *75* (19), 3473–3476. https://doi.org/10.1103/PhysRevLett.75.3473.

(8) Ross, M.; McMahan, A. K. Systematics of the s → d and p → d Electronic Transition at High Pressure for the Elements I through La. *Phys. Rev. B* **1982**, *26* (8), 4088–4093. https://doi.org/10.1103/PhysRevB.26.4088.

(9) Dolmatov, V. K.; Baltenkov, A. S.; Connerade, J.-P.; Manson, S. T. Structure and Photoionization of Confined Atoms. *Radiation Physics and Chemistry* **2004**, *70* (1–3), 417–433. https://doi.org/10.1016/j.radphyschem.2003.12.024.

(10) Woolman, G.; Naden Robinson, V.; Marqués, M.; Loa, I.; Ackland, G. J.; Hermann, A. Structural and Electronic Properties of the Alkali Metal Incommensurate Phases. *Phys. Rev. Materials* **2018**, *2* (5), 053604. https://doi.org/10.1103/PhysRevMaterials.2.053604.

(11) Connerade, J. P.; Dolmatov, V. K.; Lakshmi, P. A. The Filling of Shells in Compressed Atoms. *J. Phys. B: At. Mol. Opt. Phys.* **2000**, *33* (2), 251–264. https://doi.org/10.1088/0953-4075/33/2/310.

(12) Parker, L. J.; Atou, T.; Badding, J. V. Transition Element-Like Chemistry for Potassium Under Pressure. *Science* **1996**, *273* (5271), 95–97. https://doi.org/10.1126/science.273.5271.95.

(13) Lv, Y.; Li, J.; Zhang, Z.; Geng, Y.; Xu, Z.; Liu, Y.; Yuan, J.; Wang, Q.; Wang, X. Reverse Charge Transfer and Decomposition in Ca-Te Compounds under High Pressure. *Physical Chemistry Chemical Physics* **2024**, *26*, 10399–10407. https://doi.org/10.1039/d3cp06209k.

(14) Ma, Y.; Eremets, M.; Oganov, A. R.; Xie, Y.; Trojan, I.; Medvedev, S.; Lyakhov, A. O.; Valle, M.; Prakapenka, V. Transparent Dense Sodium. *Nature* **2009**, *458* (7235), 182–185. https://doi.org/10.1038/nature07786.

(15) Oganov, A. R.; Ma, Y.; Xu, Y.; Errea, I.; Bergara, A.; Lyakhov, A. O. Exotic Behavior and Crystal Structures of Calcium under Pressure. *Proc. Natl. Acad. Sci. U.S.A.* **2010**, *107* (17), 7646–7651. https://doi.org/10.1073/pnas.0910335107.

(16) Olijnyk, H.; Holzapfel, W. B. Phase Transitions in Alkaline Earth Metals under Pressure. *Physics Letters A* **1984**, *100* (4), 191–194. https://doi.org/10.1016/0375-9601(84)90757-6.

(17) Yabuuchi, T.; Nakamoto, Y.; Shimizu, K.; Kikegawa, T. New High-Pressure Phase of Calcium.

*J. Phys. Soc. Jpn.* **2005**, *74* (9), 2391–2392. https://doi.org/10.1143/JPSJ.74.2391.

(18) McMahon, M. I.; Bovornratanaraks, T.; Allan, D. R.; Belmonte, S. A.; Nelmes, R. J. Observation of the Incommensurate Barium-IV Structure in Strontium Phase V. *Phys. Rev. B* **2000**, *61* (5), 3135–3138. https://doi.org/10.1103/PhysRevB.61.3135.

(19) Arapan, S.; Mao, H.; Ahuja, R. Prediction of Incommensurate Crystal Structure in Ca at High Pressure. *Proc. Natl. Acad. Sci. U.S.A.* **2008**, *105* (52), 20627–20630. https://doi.org/10.1073/pnas.0810813105.

(20) Yabuuchi, T.; Matsuoka, T.; Nakamoto, Y.; Shimizu, K. Superconductivity of Ca Exceeding 25 K at Megabar Pressures. *J. Phys. Soc. Jpn.* **2006**, *75* (8), 083703. https://doi.org/10.1143/JPSJ.75.083703.

(21) McMullen, J. S.; Huo, R.; Vasko, P.; Edwards, A. J.; Hicks, J. Anionic Magnesium and Calcium Hydrides: Transforming CO into Unsaturated Disilyl Ethers. *Angew Chem Int Ed* **2023**, *62* (1), e202215218. https://doi.org/10.1002/anie.202215218.

(22) Zhang, L.; Shi, G.; Peng, B.; Gao, P.; Chen, L.; Zhong, N.; Mu, L.; Zhang, L.; Zhang, P.; Gou, L.; Zhao, Y.; Liang, S.; Jiang, J.; Zhang, Z.; Ren, H.; Lei, X.; Yi, R.; Qiu, Y.; Zhang, Y.; Liu, X.; Wu, M.; Yan, L.; Duan, C.; Zhang, S.; Fang, H. Novel 2D CaCl Crystals with Metallicity, Room-Temperature Ferromagnetism, Heterojunction, Piezoelectricity-like Property and Monovalent Calcium Ions. *National Science Review* **2020**, nwaa274. https://doi.org/10.1093/nsr/nwaa274.

(23) Kong, J.; Shi, K.; Oganov, A. R.; Zhang, J.; Su, L.; Dong, X. Exotic Compounds of Monovalent Calcium Synthesized at High Pressure. *Matter and Radiation at Extremes* **2024**, *9* (6), 067803. https://doi.org/10.1063/5.0222230.

(24) Li, Y.-L.; Wang, S.-N.; Oganov, A. R.; Gou, H.; Smith, J. S.; Strobel, T. A. Investigation of Exotic Stable Calcium Carbides Using Theory and Experiment. *Nat Commun* **2015**, *6* (1), 6974. https://doi.org/10.1038/ncomms7974.

(25) Dong, X.; Oganov, A. R.; Cui, H.; Zhou, X.-F.; Wang, H.-T. Electronegativity and Chemical Hardness of Elements under Pressure. *Proc. Natl. Acad. Sci. U.S.A.* **2022**, *119* (10), e2117416119. https://doi.org/10.1073/pnas.2117416119.

(26) Sanville, E.; Kenny, S. D.; Smith, R.; Henkelman, G. Improved Grid-based Algorithm for Bader Charge Allocation. *J Comput Chem* **2007**, *28* (5), 899–908. https://doi.org/10.1002/jcc.20575.

(27) Henkelman, G.; Arnaldsson, A.; Jónsson, H. A Fast and Robust Algorithm for Bader Decomposition of Charge Density. *Computational Materials Science* **2006**, *36* (3), 354–360. https://doi.org/10.1016/j.commatsci.2005.04.010.

(28) Tang, W.; Sanville, E.; Henkelman, G. A Grid-Based Bader Analysis Algorithm without Lattice Bias. *Journal of Physics Condensed Matter* **2009**, *21* (8). https://doi.org/10.1088/0953-8984/21/8/084204.

(29) Profeta, M.; Benoit, M.; Mauri, F.; Pickard, C. J. First-Principles Calculation of the $^{17}$O NMR Parameters in Ca Oxide and Ca Aluminosilicates: The Partially Covalent Nature of the Ca−O Bond, a Challenge for Density Functional Theory. *J. Am. Chem. Soc.* **2004**, *126* (39), 12628–12635. https://doi.org/10.1021/ja0490830.

(30) Posternak, M.; Baldereschi, A.; Krakauer, H.; Resta, R. Non-Nominal Value of the Dynamical Effective Charge in Alkaline-Earth Oxides. *Phys. Rev. B* **1997**, *55* (24), R15983–R15986. https://doi.org/10.1103/PhysRevB.55.R15983.

(31) Sun, H.; Deng, K.; Kan, E.; Du, Y. Second-Order Jahn–Teller Effect Induced High-Temperature Ferroelectricity in Two-Dimensional $NbO_2$ X (X = I, Br). *Nanoscale Adv.* **2023**, *5* (11), 2979–2985. https://doi.org/10.1039/D3NA00245D.

(32) Yao, Q.-F.; Cai, J.; Tong, W.-Y.; Gong, S.-J.; Wang, J.-Q.; Wan, X.; Duan, C.-G.; Chu, J. H. Manipulation of the Large Rashba Spin Splitting in Polar Two-Dimensional Transition-Metal Dichalcogenides. *Phys. Rev. B* **2017**, *95* (16), 165401. https://doi.org/10.1103/PhysRevB.95.165401.

(33) Karaziya, R. I. Excited Electron Orbit Collapse and Atomic Spectra. *Sov. Phys. Usp.* **1981**, *24* (9), 775–794. https://doi.org/10.1070/PU1981v024n09ABEH004823.

(34) Peng, F.; Wang, Y.; Wang, H.; Zhang, Y.; Ma, Y. Stable Xenon Nitride at High Pressures. *Phys. Rev. B* **2015**, *92* (9), 094104. https://doi.org/10.1103/PhysRevB.92.094104.

(35) Li, J.; Geng, Y.; Xu, Z.; Zhang, P.; Garbarino, G.; Miao, M.; Hu, Q.; Wang, X. Mechanochemistry and the Evolution of Ionic Bonds in Dense Silver Iodide. *Journal of the American Chemical Society* **2022**. https://doi.org/10.1021/jacsau.2c00550.

(36) Liu, Y.; Li, J.; Geng, Y.; Xu, Z.; Lv, Y.; Zhang, Z.; Yuan, J.; Wang, X. Pressure-Induced Phase Transitions and Decompositions of Sr–S Compounds. *Physica B: Condensed Matter* **2024**, *681*, 415846. https://doi.org/10.1016/j.physb.2024.415846.

(37) Xu, Z.; Li, J.; Geng, Y.; Zhang, Z.; Lv, Y.; Zhang, C.; Wang, Q.; Wang, X. Regulation of Ionic Bond in Group IIB Transition Metal Iodides. *Chinese Physics Letters* **2023**, *40* (7), 076201. https://doi.org/10.1088/0256-307X/40/7/076201.

(38) Xu, Z.; Rui, Q.; Geng, Y.; Li, J.; Wang, Q.; Wang, X. Pressure-Induced Decomposition of Cadmium Iodide. *EPL* **2022**, *140* (1), 16003. https://doi.org/10.1209/0295-5075/ac94f4.

(39) Petkov, V.; Zafar, A.; Kenesei, P.; Shastri, S. Chemical Compression and Ferroic Orders in La Substituted BiFe O 3. *Phys. Rev. Materials* **2023**, *7* (5), 054404. https://doi.org/10.1103/PhysRevMaterials.7.054404.

(40) Ibrahim, A.; Paskevicius, M.; Buckley, C. E. Chemical Compression and Transport of Hydrogen Using Sodium Borohydride. *Sustainable Energy Fuels* **2023**, *7* (5), 1196–1203. https://doi.org/10.1039/D2SE01334G.

(41) Wang, X.; Li, J.; Khodagholian, D.; Lin, J.; Jackson, M.; Spera, F.; Redfern, S.; Miao, M. Reverse Chemistry of Iron under High Pressure and the Distribution of Elements in the Deep Earth. **2020**. https://doi.org/10.21203/rs.3.rs-130002/v1.

(42) Kennedy, J.; Eberhart, R. Particle Swarm Optimization. In *Proceedings of ICNN'95 - International Conference on Neural Networks*; IEEE: Perth, WA, Australia, 1995; Vol. 4, pp 1942–1948. https://doi.org/10.1109/ICNN.1995.488968.

(43) Eberhart, R.; Kennedy, J. A New Optimizer Using Particle Swarm Theory. In *MHS'95. Proceedings of the Sixth International Symposium on Micro Machine and Human Science*; IEEE: Nagoya, Japan, 1995; pp 39–43. https://doi.org/10.1109/MHS.1995.494215.

(44) Wang, Y.; Lv, J.; Zhu, L.; Ma, Y. Crystal Structure Prediction via Particle-Swarm Optimization. *Physical Review B - Condensed Matter and Materials Physics* **2010**, *82* (9). https://doi.org/10.1103/PhysRevB.82.094116.

(45) Wang, Y.; Lv, J.; Zhu, L.; Ma, Y. CALYPSO: A Method for Crystal Structure Prediction. *Computer Physics Communications* **2012**, *183* (10), 2063–2070. https://doi.org/10.1016/j.cpc.2012.05.008.

(46) Wang, X.; Wang, Y.; Miao, M.; Zhong, X.; Lv, J.; Cui, T.; Li, J.; Chen, L.; Pickard, C. J.; Ma,

Y. Cagelike Diamondoid Nitrogen at High Pressures. *Physical Review Letters* **2012**, *109* (17). https://doi.org/10.1103/PhysRevLett.109.175502.

(47) Perdew, J. P.; Burke, K.; Ernzerhof, M. Generalized Gradient Approximation Made Simple. *Phys. Rev. Lett.* **1996**, *77* (18), 3865–3868. https://doi.org/10.1103/PhysRevLett.77.3865.

(48) Hafner, J. Ab-initio Simulations of Materials Using VASP: Density-functional Theory and Beyond. *J Comput Chem* **2008**, *29* (13), 2044–2078. https://doi.org/10.1002/jcc.21057.

(49) Sun, G.; Kürti, J.; Rajczy, P.; Kertesz, M.; Hafner, J.; Kresse, G. Performance of the Vienna Ab Initio Simulation Package (VASP) in Chemical Applications. *Journal of Molecular Structure: THEOCHEM* **2003**, *624* (1–3), 37–45. https://doi.org/10.1016/S0166-1280(02)00733-9.

(50) Togo, A.; Tanaka, I. First Principles Phonon Calculations in Materials Science. *Scripta Materialia* **2015**, *108*, 1–5. https://doi.org/10.1016/j.scriptamat.2015.07.021.

(51) Deringer, V. L.; Tchougréeff, A. L.; Dronskowski, R. Crystal Orbital Hamilton Population (COHP) Analysis as Projected from Plane-Wave Basis Sets. *Journal of Physical Chemistry A* **2011**, *115* (21), 5461–5466. https://doi.org/10.1021/jp202489s.

(52) Maintz, S.; Deringer, V. L.; Tchougréeff, A. L.; Dronskowski, R. LOBSTER: A Tool to Extract Chemical Bonding from Plane-Wave Based DFT. *Journal of Computational Chemistry* **2016**, *37* (11), 1030–1035. https://doi.org/10.1002/jcc.24300.

(53) Nelson, R.; Ertural, C.; George, J.; Deringer, V. L.; Hautier, G.; Dronskowski, R. LOBSTER : Local Orbital Projections, Atomic Charges, and Chemical-bonding Analysis from PROJECTOR-AUGMENTED-WAVE-BASED Density-functional Theory. *J Comput Chem* **2020**, *41* (21), 1931–1940. https://doi.org/10.1002/jcc.26353.

(54) Zhuang, Y.; Li, J.; Lu, W.; Yang, X.; Du, Z.; Hu, Q. High Temperature Melting Curve of Basaltic Glass by Laser Flash Heating. *Chinese Phys. Lett.* **2022**, *39* (2), 020701. https://doi.org/10.1088/0256-307X/39/2/020701.

(55) Du, Z.; Amulele, G.; Robin Benedetti, L.; Lee, K. K. M. Mapping Temperatures and Temperature Gradients during Flash Heating in a Diamond-Anvil Cell. *Review of Scientific Instruments* **2013**, *84* (7), 075111. https://doi.org/10.1063/1.4813704.

(56) Mao, H. K.; Xu, J.; Bell, P. M. Calibration of the Ruby Pressure Gauge to 800 Kbar under Quasi-hydrostatic Conditions. *J. Geophys. Res.* **1986**, *91* (B5), 4673–4676. https://doi.org/10.1029/JB091iB05p04673.

(57) Akahama, Y.; Kawamura, H. Pressure Calibration of Diamond Anvil Raman Gauge to 310GPa. *Journal of Applied Physics* **2006**, *100* (4), 043516. https://doi.org/10.1063/1.2335683.

(58) Holland, T. J. B.; Redfern, S. A. T. Unit Cell Refinement from Powder Diffraction Data: The Use of Regression Diagnostics. *Mineral. mag.* **1997**, *61* (404), 65–77. https://doi.org/10.1180/minmag.1997.061.404.07.

## Figures

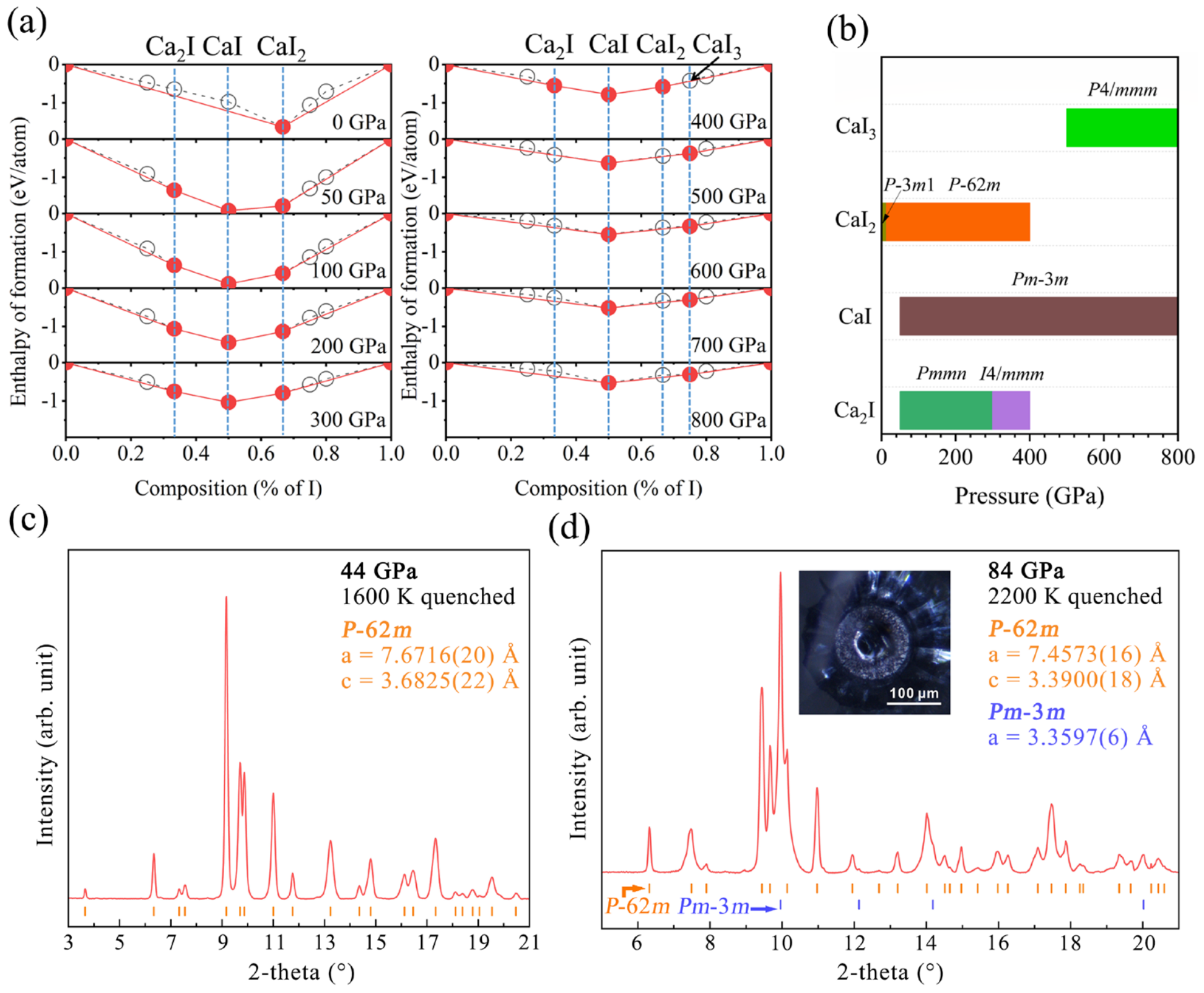

**Figure 1.** Ca-I compounds under high pressure. (a) Convex hull diagrams for the Ca-I system from ambient to 800 GPa. The compounds located on the convex hull (solid red circles) are stable against decomposition into other compositions. (b) Stability fields of various Ca-I compounds. (c) XRD pattern of $CaI_2$ at 44 GPa and quenched from 1600 K. Orange ticks below the diffraction pattern represent the Bragg positions of hexagonal $CaI_2$. Data were acquired using an X-ray wavelength of 0.4246 Å. (d) XRD pattern at 84 GPa and quenched from 2200 K, showing the coexistence of hexagonal $CaI_2$ and the *Pm*3*m* CaI (blue ticks). Data were acquired using an X-ray wavelength of 0.4127 Å. Inset is a photograph of $CaI_2$ loaded in the DAC sample chamber.

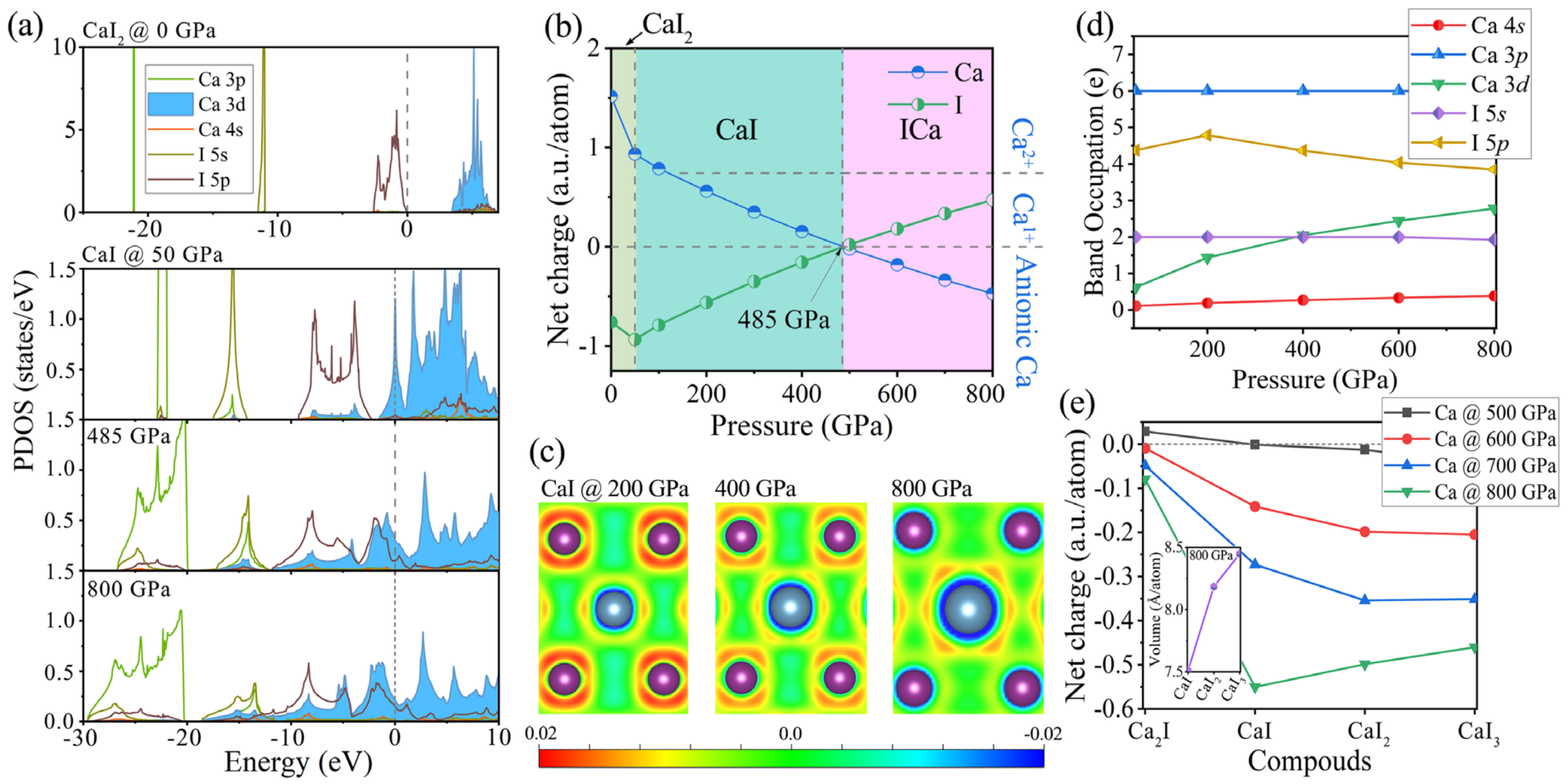

**Figure 2.** Electronic properties. (a) PDOS of Ca-I Compounds at a given pressure. (b) The net charge of Ca and I atoms in Ca-I compounds is based on Bader analysis. (c) Plotted the difference in charge density of the Ca-I compounds at a given pressure. (d) Band occupancy numbers of CaI in *Pm*-3*m* at 50-800 GPa. (e) The net charge of Ca and I atoms in different stoichiometry compounds is based on Bader analysis.

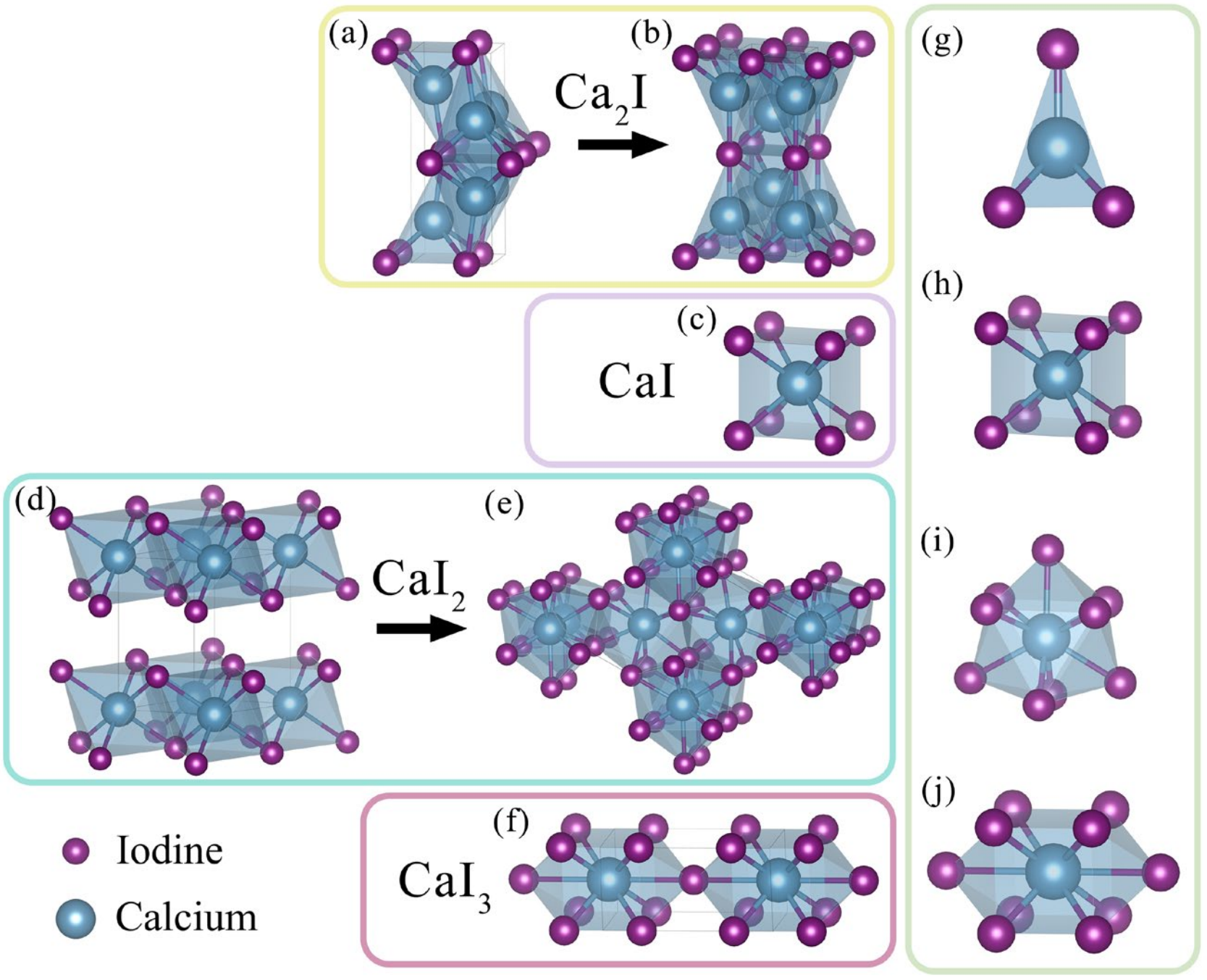

**Figure 3.** Crystal structures of Ca-I compounds. (a) $Ca_2I$ in *Pmmn* structure. (b) $Ca_2I$ in *I*4/*mmm* structure. (c) CaI in *Pm*-3*m* structure. (d) $CaI_2$ in *P*-3*m*1 structure. (e) $CaI_2$ in *P*-62*m* structure. (f) $CaI_3$ in *P*4/*mmm* structure. (g) - (j) The units that makeup structure $Ca_2I$ in *I*4/*mmm*, CaI in *Pm*-3*m*, $CaI_2$ in *P*-3*m*1, and $CaI_3$ in *P*4/*mmm*.

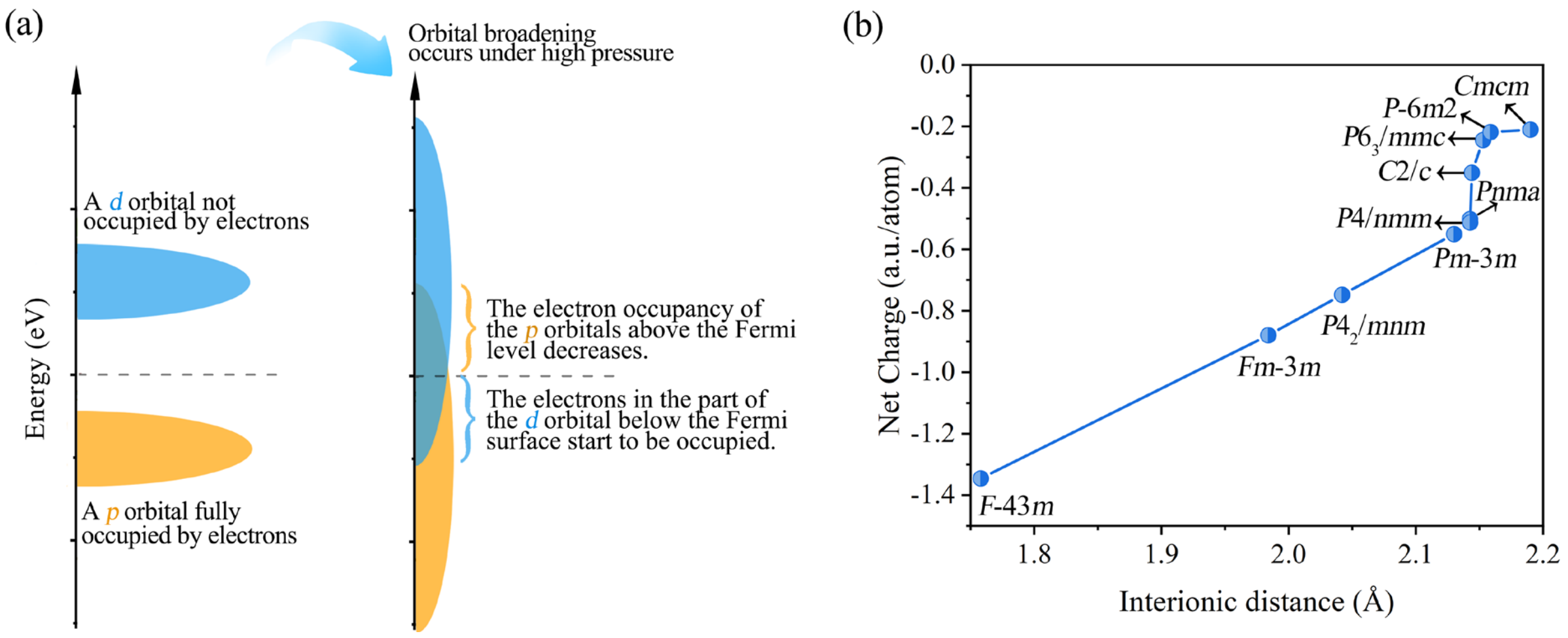

**Figure 4.** The orbital broadening of the CaI. (a) Schematic diagram of the orbital broadening phenomenon. (b) Ca-I interionic distance versus net Charge of various CaI crystal structures at 800 GPa, in which the B2 phase (*Pm*-3*m*) is the ground state.

# Unlock Anionic Behavior of Calcium Through Pressure Engineering

Yang Lv,[1] Junwei Li,[2] Jianfu Li,[1,*] Yong Liu,[1] Jianan Yuan,[1] Jiani Lin,[1] Saori Kawaguchi-Imada[3], Qingyang Hu,[2,*] and Xiaoli Wang[1,*]

[1]*School of Physics and Electronic Information, Yantai University, Yantai 264005, China*

[2]*Center for High Pressure Science and Technology Advanced Research, Beijing 100094, China*

[3]*Japan Synchrotron Radiation Research Institute, Hyogo, Japan*

Corresponding authors: jianfuli@ytu.edu.cn (Jianfu Li); qingyang.hu@hpstar.ac.cn (Qingyang Hu); xlwang@ytu.edu.cn (Xiaoli wang)

**Supplementary Note 1: Experimental detail**

In experiment Run 1(Figure 1c), the starting material was loaded in a 150 μm diameter hole in a rhenium gasket intended by diamond anvils with 250 μm diameter culets. A ruby ball was loaded peripherally to the sample on the cylinder side and served as a pressure calibrant. The sample was compressed up to ~42 GPa and then heated by an ND: YAG laser with a laser-heating spot of ~30 μm at ~1600 K in the central area. The temperature-quenched sample was examined in situ at high pressures by synchrotron X-ray powder diffraction on the BL04 beamline at ALBA. The X-ray beam size was 15×18 $\mu m^2$ with a wavelength of 0.4246 Å. The unit cell is calculated by indexing the 100, $2\bar{1}0$, 200, 101, $2\bar{1}1$, $3\bar{1}0$, 201, 300, and $3\bar{1}1$ diffraction peak of hexagonal $CaI_2$ by a nonlinear fitting program.

In experiment Run 2(Figure 1d), the starting material was loaded in a 60 μm diameter hole in a rhenium gasket intended by diamond anvils with 150 μm diameter culets beveled from 300 μm culets. The calibrated first-order Raman band shift of diamond anvils was used as a pressure gauge. The sample was compressed up to ~94 GPa and then heated to ~2200 K in the flat top area created by a ~20 μm diameter laser-heating spot. The sample was subsequently quenched to room temperature and examined in situ at high pressures by synchrotron X-ray powder diffraction at beamline BL10XU of SPring-8. The X-ray beam size was 10×10 $\mu m^2$ with a wavelength of 0.4127 Å. The unit cell is calculated by indexing the $2\bar{1}0$, 200, 101, $2\bar{1}1$, $3\bar{1}0$, 201, 300, $3\bar{1}1$, $4\bar{1}0$ diffraction peaks of hexagonal $CaI_2$ and 110, 111, 200, 220 diffraction peaks of cubic $CaI_2$, respectively.

**Supplementary Note 2: Hyper-coordination in high-pressure Ca-I compounds**

Specifically, the structural evolution of the $CaI_2$ compound with pressure is notable. At ambient pressure, the most stable phase of $CaI_2$ crystallizes in the *P*-3*m*1 space group, forming a layered compound composed of octahedral units (as shown in Figure 3(d) and (e)). In this structure, each calcium atom is coordinated by six iodine atoms, with the shortest Ca-I bond length measured at 3.181 Å. Upon compression to 44 GPa, a phase transition occurs to the *P*-62*m* high-pressure phase. Figure 3(i) illustrates that this high-pressure phase adopts a denser structure characterized by nine-coordinate tri-capped hexagonal prism units. These structural units can be visualized as three pyramidal caps protruding from the face centers of a hexagonal prism, resulting in a significantly more compact arrangement. The shortest bond length in this phase decreases to 2.755 Å, reflecting the enhanced bonding interactions. Subsequently, at 50 GPa, the CaI compound with the *Pm*-3*m* space group fell to the bottom of the convex hull, as shown in Figure 3(c). The *Pm*-3*m* structure represents a classical structure in binary ionic compounds, where each Ca atom is coordinated with eight I atoms to form a cubic unit cell with a bond length of 2.881 Å. This structure remains stable throughout the entire pressure range investigated in this study. Another intriguing evolution under pressure is the phase transition of $Ca_2I$ from the *Pmmn* to the *I*4/*mmm* space group, as demonstrated in Figures 3(a) and (b). The *Pmmn* phase is constructed by stacking distorted square pyramids, where at 50 GPa, the bond lengths between Ca and the bottom I atoms are 2.855 Å, while those between Ca and the top I atoms are 2.942 Å. Upon compression beyond 300 GPa, the system undergoes a structural transformation to the *I*4/*mmm* phase. In this new phase, the distorted square pyramids are replaced by regular ones, resulting in bond lengths of 2.440 Å between Ca and the four bottom I atoms, and 2.462 Å between Ca and the top I atom. Finally, the most iodine-rich structure among all stable Ca-I compounds is $CaI_3$ with the *P*4*mmm* space group, as illustrated in Figure 3(f). This structure exhibits the highest coordination number among the investigated systems, with each Ca atom coordinated to ten I atoms. The structural units adopt a double-capped cube configuration, which can be visualized as an octahedron formed by eight I atoms (with bond lengths of 2.262 Å) surrounding a central Ca atom, with two additional I atoms (bond lengths of 2.794 Å) attached to two opposite faces, forming a dodecahedral geometry.

Figure 2(e) is the Bader charges of the high-pressure phase for the aforementioned chemical compositions as a function of iodine content. The results reveal a correlation between the electronic properties and structural selectivity of the compounds under pressure. At 500 GPa, Ca remains reduced (acting as a reducing agent) in the Ca-rich compounds, while in the I-rich compounds, calcium begins to exhibit oxidizing behavior by losing electrons. This trend becomes more pronounced at 600 and 700 GPa, with the curves shifting obviously toward the I-rich side, indicating a higher electron-donating efficiency of calcium in the I-rich compounds than in the Ca-rich side. It can also be seen from Figure S1 that the pressure for the metal-to-nonmetal transition of Ca gradually decreases from the Ca-rich side to the I-rich side. However, this trend deviates at 800 GPa, where Ca in the Ca-rich side exhibits lower electron-accepting efficiency both thermodynamically (in the convex hull where $Ca_2I$ becomes metastable) and electronically. Interestingly, the curve shifts downward toward the CaI (*Pm*-3*m*), suggesting enhanced electron-accepting capability of Ca in CaI. This anomaly can be attributed to the unit-cell volumes at 800 GPa, as shown in Figure 2(e). The volume data reveal that CaI has a smaller atomic volume (7.55 $Å^3$) compared to $CaI_2$ (8.21 $Å^3$) and $CaI_3$ (8.45 $Å^3$). This reduced unit-cell size introduces a "chemical compression" effect, which enhances the electron-accepting ability of Ca in CaI, enabling it to acquire more electrons. In addition, the conclusion above can also account for the phenomenon in Figure 1(a), where the structure at the bottom of the convex hull line first shifts to the left and then to the right.

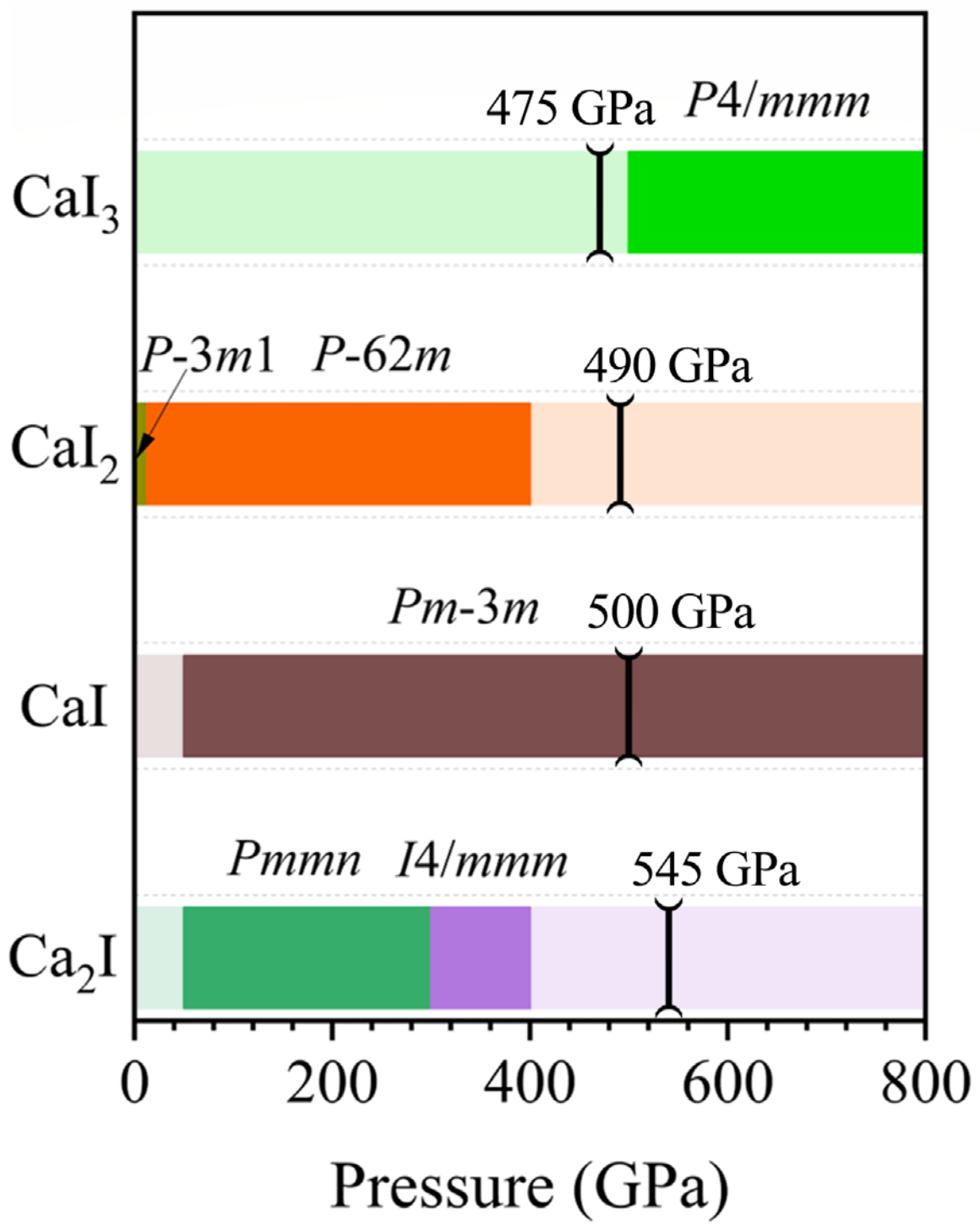

**Figure S1.** The pressure for the metal-nonmetal transition of Ca in Ca-I compounds under different ratios

**Supplementary Note 3: Origin of the stability of Ca-I compounds**

To understand the anionic behavior of Ca in a variety of high-pressure calcium iodides, we continue to extract the lowest $\Delta H$ of the Ca-I structures. A phenomenon is that the $\Delta H$ of Ca-I compounds decreases with increasing pressure in the pressure range 0-80 GPa, but within 80-700 GPa, the $\Delta H$ increases, while between 700-1000 GPa, the $\Delta H$ decreases again, as shown in Figure S2(a). This atypical tendency indicates that the stability of Ca-I compounds is not simply increasing or decreasing with increasing pressure and reveals the complexity of pressure on the stability of compounds.

To understand the atypical tendency of $\Delta H$, the $\Delta H$ of the Ca-I compound and its energy components, including the difference of internal energy ($\Delta U$) and the $\Delta PV$ term, are calculated, as shown in Figure S2(b). At lower pressure (< 80 GPa), the $\Delta H$ and its decrease are mainly caused by $\Delta U$. In the pressure range of 80-700 GPa, the $\Delta H$ increases as the pressure rises. Within this range, the increase of $\Delta H$ is caused by the dramatic increase of $\Delta PV$ in the pressure range of 80-300 GPa, although the $\Delta H$ is mainly composed of the $\Delta U$. In the pressure range of 300-700 GPa, the $\Delta PV$ has a positive value, the $\Delta H$, and its increase is mainly contributed by $\Delta U$. Above 700 GPa, from Figure S2(c), the volume of the compound under pressure becomes smaller than that of the corresponding elemental mixture. At this point, the value of $\Delta PV$ becomes negative again, while the $\Delta U$ is still increasing and turns positive at 975 GPa. In this pressure range (700-1000 GPa), the decrease of $\Delta H$ is caused by the $\Delta PV$. It should be noted that beyond 825 GPa, the change in enthalpy ($\Delta H$) is predominantly attributed to the $\Delta PV$ term. When pressure is below 700 GPa, the increased internal energy is the main reason that causes the weakening of stability, except in the pressure range of 80-300 GPa. Above 700 GPa, the $\Delta PV$ term is the key fact. Especially, above 825 GPa, the re-stability of the Ca-I compound is mainly contributed to by the $\Delta PV$ term. The calculated volume difference ($\Delta V$) between the Ca-I compounds and Ca + I, as shown in Figure S2(c), indicates that the Ca-I compound (CaI in *Pm*-3*m*) has a smaller average volume than elemental members after 700 GPa.

Furthermore, by comparing the trend of Bader charge in Figure 2(b) with $\Delta H$, we can also find that, after excluding the covalent interactions in Figure S4, the ionic interactions also significantly impact the system's thermodynamic stability.

The influence of pressure on the thermodynamic stability of compounds is complex and multi-dimensional, but interesting. Pressure can make compounds more stable, such as enabling the formation of compounds from noble gases[1]. Pressure can also make compounds unstable and decompose, such as AgI[2], SrS[3], and Group IIB iodide compounds[4,5]. However, the trend of first becoming stable, then decreasing instability, and then becoming stable again, as shown in Figure 1c, is currently a very rare phenomenon. The metal-to-nonmetal transition and the atypical stability trend of compounds under pressure have inevitably led us to rethink the rules of physical property changes of substances under pressure, which is precisely one of the motivations for researchers to conduct more in-depth exploration of the effects of pressure.

As shown in Figure S2(b), at higher pressures (>700 GPa), a low $\Delta PV$ term is the primary factor maintaining compounds' thermodynamic stability. From the perspective of thermodynamic stability, although phases such as *F*-43*m*, *Fm*-3*m*, and $P4_2/nmm$ exhibit lower internal energy due to relatively higher charge transfer, only the B2-CaI phase with a lower $\Delta PV$ term (higher density) remains on the convex hull overall. This is corroborated by the calculated formation enthalpy per formula unit (Figure S3), where formation enthalpy increases significantly with decreasing compactness.

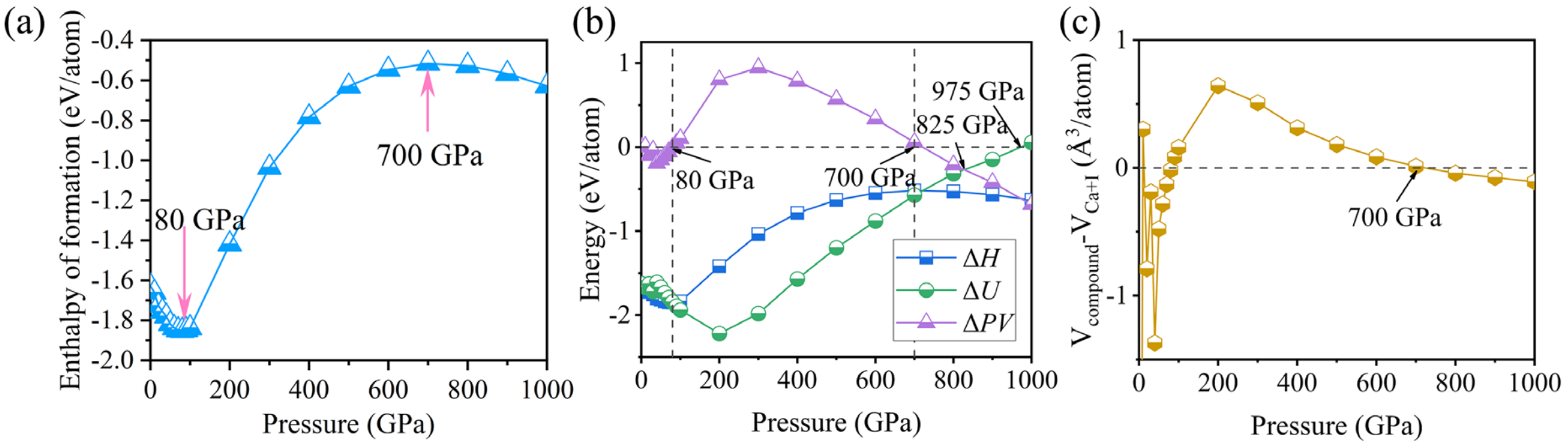

**Figure S2.** Origin of stability of Ca-I phases. (a) Enthalpies of formation of Ca-I stable phases under a series of pressures. (b) $\Delta H$, $\Delta U$, and $\Delta PV$ versus pressure for CaI. (c) The volume difference between the Ca-I compounds and the Ca + I.

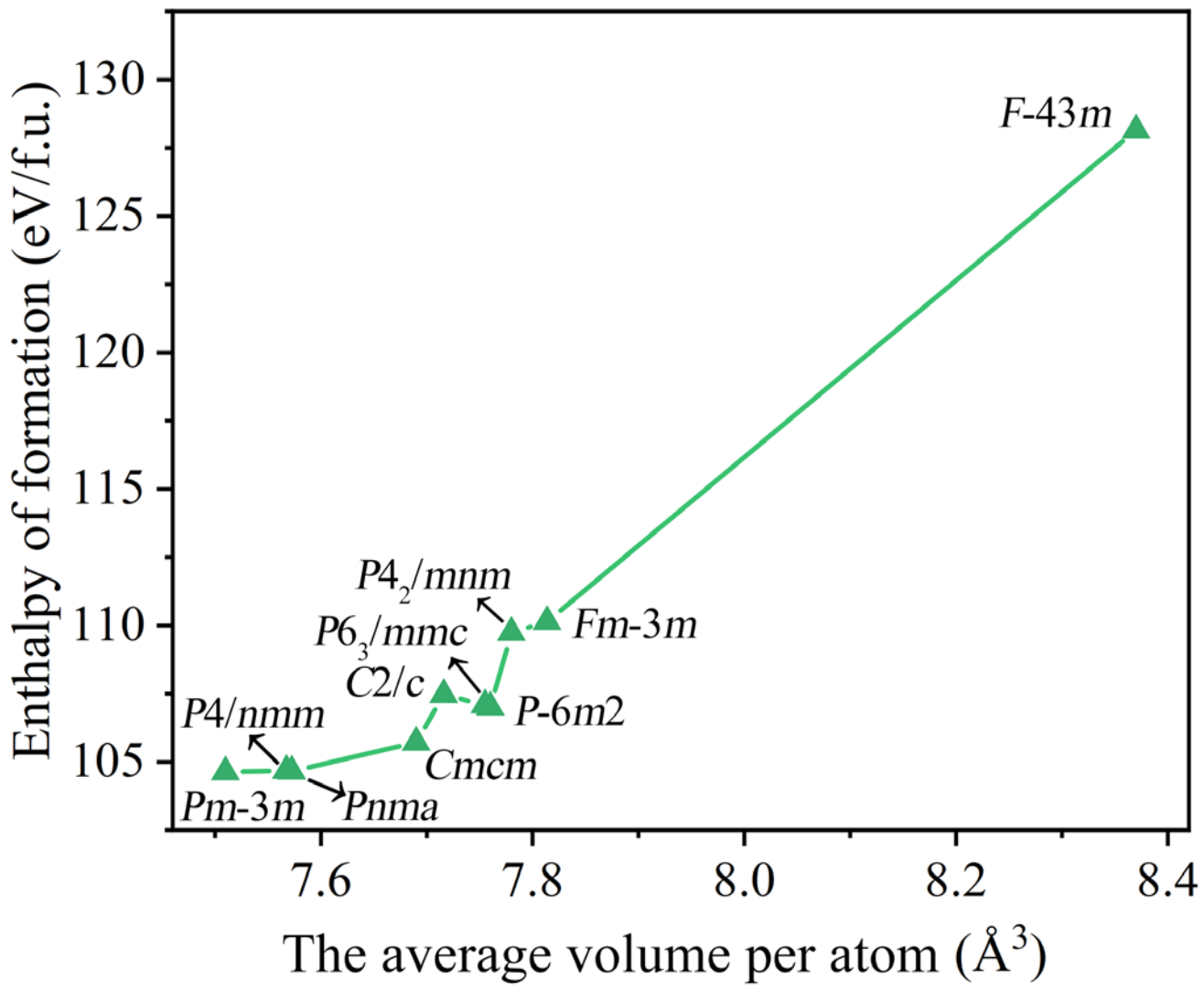

**Figure S3**. Enthalpy of different models at 800 GPa.

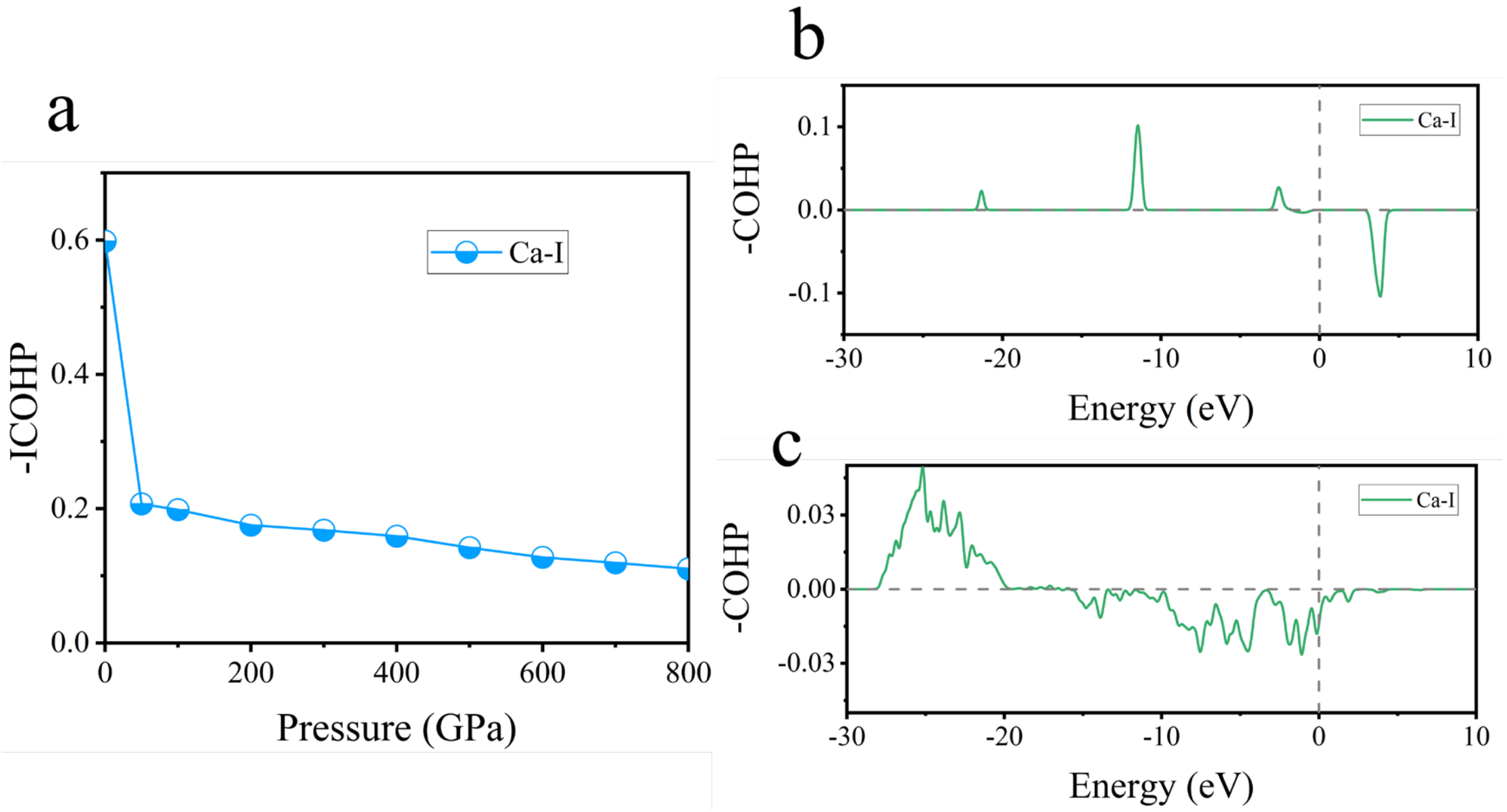

**Figure S4.** (a) Negative integrated Crystal orbital Hamilton Populations (-ICOHP) for Ca and I atoms in Ca-I compounds. The value of –ICOHP decreases with increasing pressure, which means the orbital overlap between Ca and I decreases at high pressure. (b) Calculated –COHP for $CaI_2$ in *P*-3*m*1 structure at 0 GPa. (c) Calculated –COHP for CaI in *Pm*-3*m* structure at 600 GPa.

**Supplementary Note 4:Underestimate of 3*d* orbitals of Ca**

Previous studies have shown that Ca-containing compounds fail to well reproduce experimental Nuclear Magnetic Resonance (NMR) data in PBE-DFT calculations[6,7]. This is attributed to the easy hybridization between the low-lying Ca-3*d* orbitals and orbitals of non-metallic elements, which causes deviations between calculated results and experimental values. The solution is to rigidly shift the low-lying Ca-3*d* orbitals upward by 3.2 eV[6]; in VASP calculations, we achieved this using the orbital-selective external potential (OSEP) methd[8,9]. As shown in Figure S5(a), after applying a 3.2 eV external potential, the Ca-3*d* orbitals were successfully elevated by 3.228 eV, matching the experimental reference value. Subsequent net charge calculations via the Bader method (Figure S5(b)) indicated that the anionic tendency of Ca ions remained unaffected, with only the pressure of metal-nonmetal transitions shifting upward from 485 GPa to 737 GPa.

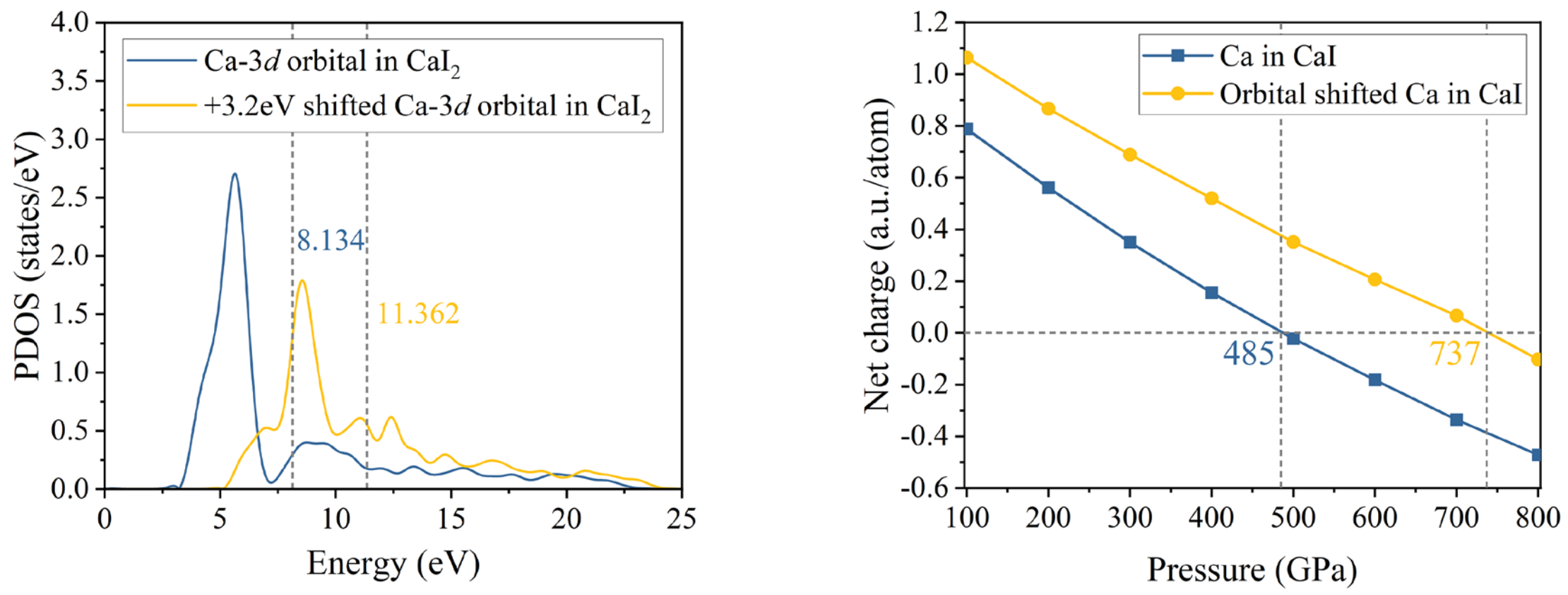

**Figure S5**. The OSEP correction for the d orbitals of Ca. (a) The DOS of the Ca-3*d* orbitals before and after OSEP correction. (b) The net charge of the Ca in CaI before and after OSEP correction.

**Table S1.** The unit-cell parameters and atomic positions of predicted structures.

| Compounds (Space group) | Pressure (GPa) | Lattice Parameter (Å, º) | Atom | Wyckoff position | *x* | *y* | *z* |
|---|---|---|---|---|---|---|---|
| $CaI_3$ (*P*4/*mmm*) | 200 | a = 2.984<br>b = 2.984<br>c = 6.042<br>α = β = γ = 90 | Ca<br>I<br>I | 1c<br>2g<br>1d | 0.5<br>0<br>0.5 | 0.5<br>0<br>0.5 | 0<br>0.228<br>0.5 |
| $CaI_2$ (*P*-3*m*1) | 10 | a = 4.045<br>b = 4.045<br>c = 6.425<br>α = β = 90<br>γ = 120 | Ca<br>I | 1a<br>2d | 0<br>0.333 | 0<br>0.667 | 0<br>0.287 |
| $CaI_2$ (*P*-62*m*) | 44 | a = 7.742<br>b = 7.742<br>c = 3.594<br>α = β = 90<br>γ = 120 | Ca<br>Ca<br>I<br>I | 2d<br>1a<br>3g<br>3f | 0.667<br>0.000<br>0.278<br>0.000 | 0.333<br>0.000<br>0.278<br>0.385 | 0.500<br>0.000<br>0.500<br>0.000 |
| CaI (*Pm*-3*m*) | 84 | a = b = c = 3.241<br>α = β = γ = 90 | Ca<br>I | 1b<br>1a | 0.5<br>0 | 0.5<br>0 | 0.5<br>0 |
| $Ca_2I$ (*Pmmn*) | 100 | a = 3.153<br>b = 8.387<br>c = 3.216<br>α = β = γ = 90 | Ca<br>I | 4e<br>2a | 0<br>0 | 0.320<br>0 | 0.196<br>0.316 |
| $Ca_2I$ (*I*4/*mmm*) | 300 | a = 2.751<br>b = 2.751<br>c = 7.931<br>α = β = γ = 90 | Ca<br>I | 4e<br>2b | 0.5<br>0 | 0.5<br>0 | 0.312<br>0.5 |
| $Ca_3I$ (*Cmcm*) | 100 | a = 5.526<br>b = 9.336<br>c = 4.284<br>α = β = γ = 90 | Ca<br>I<br>I | 8g<br>4c<br>4c | 0.776<br>0<br>0 | 0.901<br>0.180<br>0.642 | 0.250<br>0.250<br>0.250 |

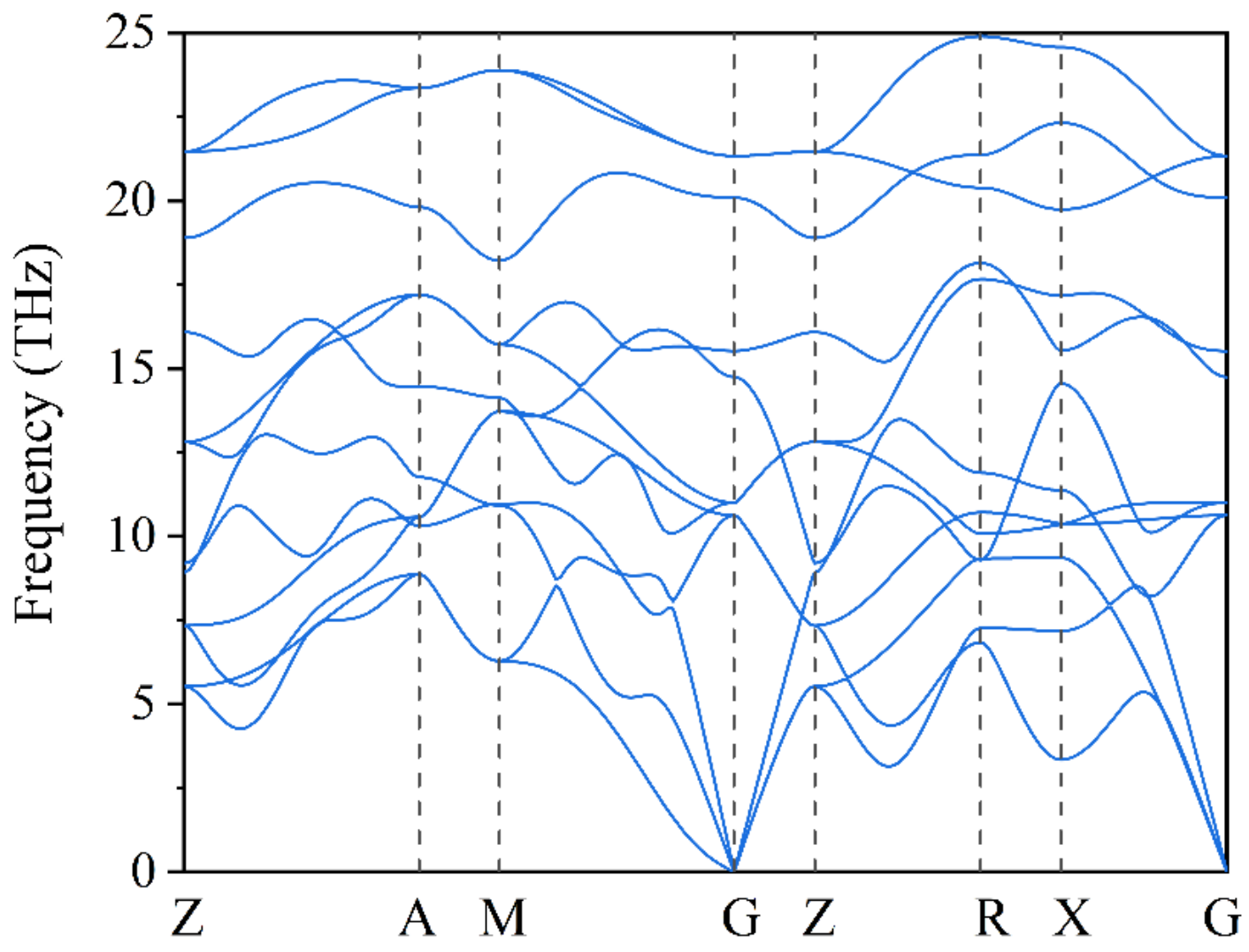

**Figure S6.** Phonon dispersion curves of $CaI_3$ in *P*4/*mmm* structure at 600 GPa.

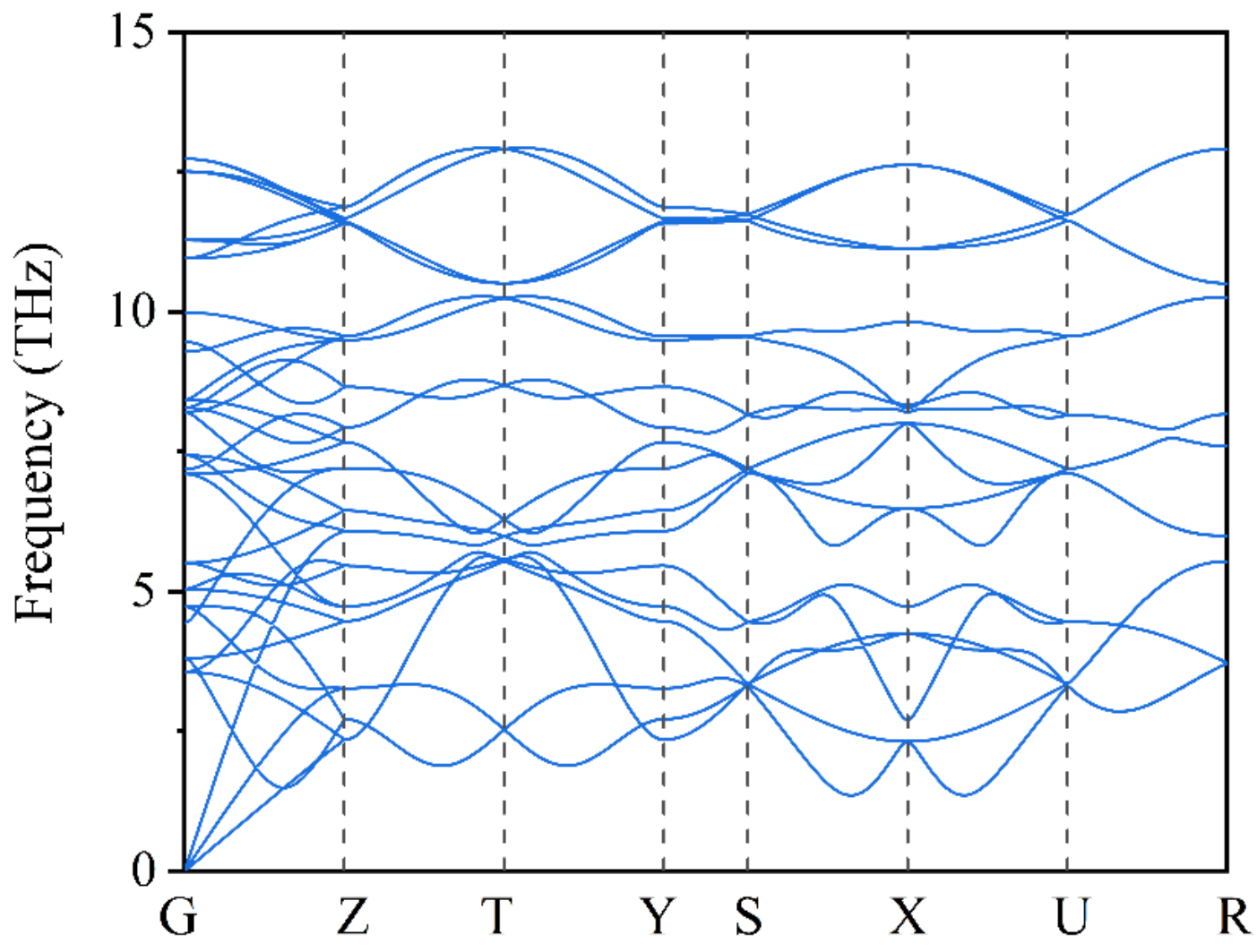

**Figure S7.** Phonon dispersion curves of $CaI_2$ in *Cmcm* structure at 200 GPa.

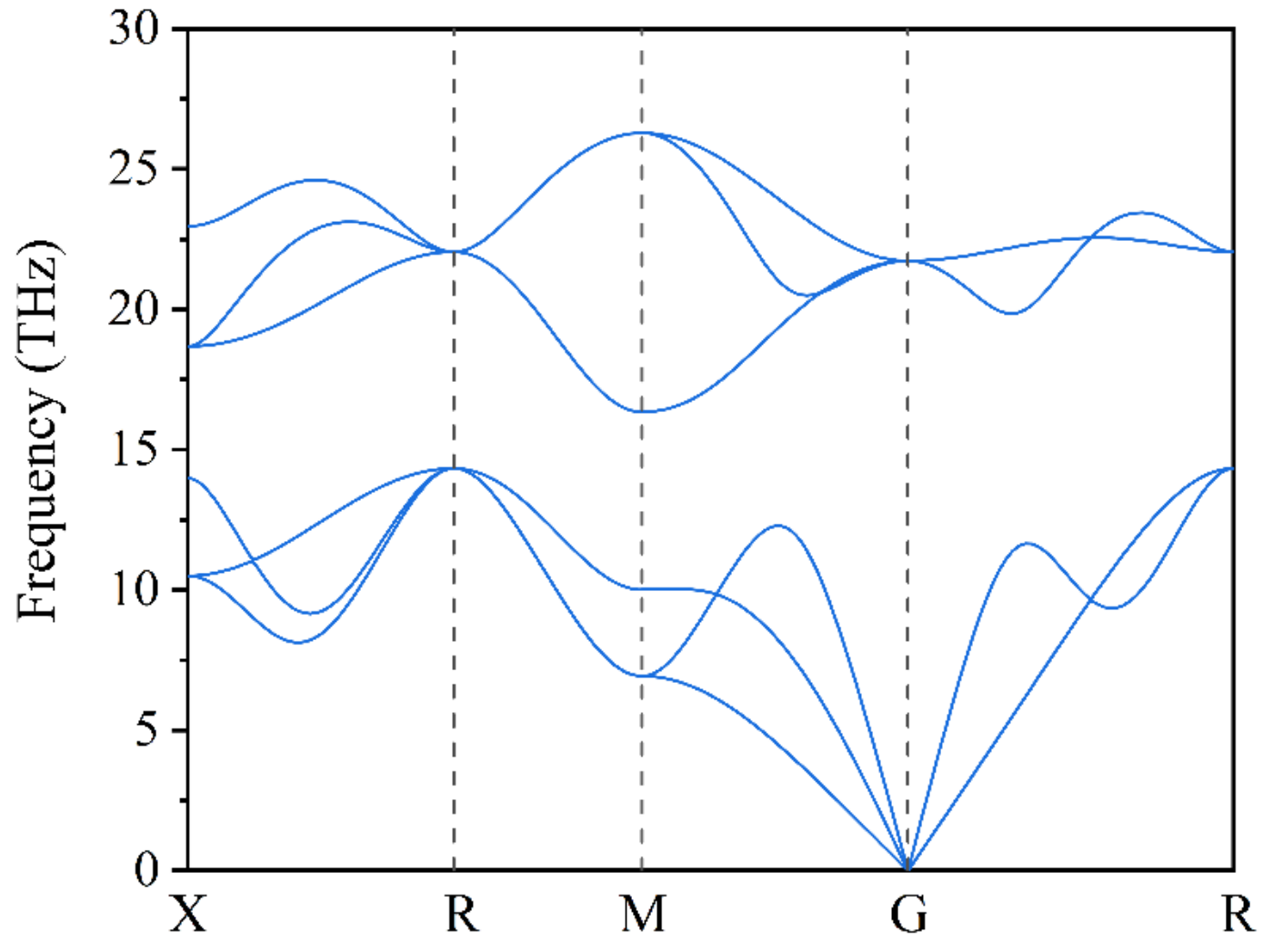

**Figure S8.** Phonon dispersion curves of CaI in *Pm*-3*m* structure at 200 GPa.

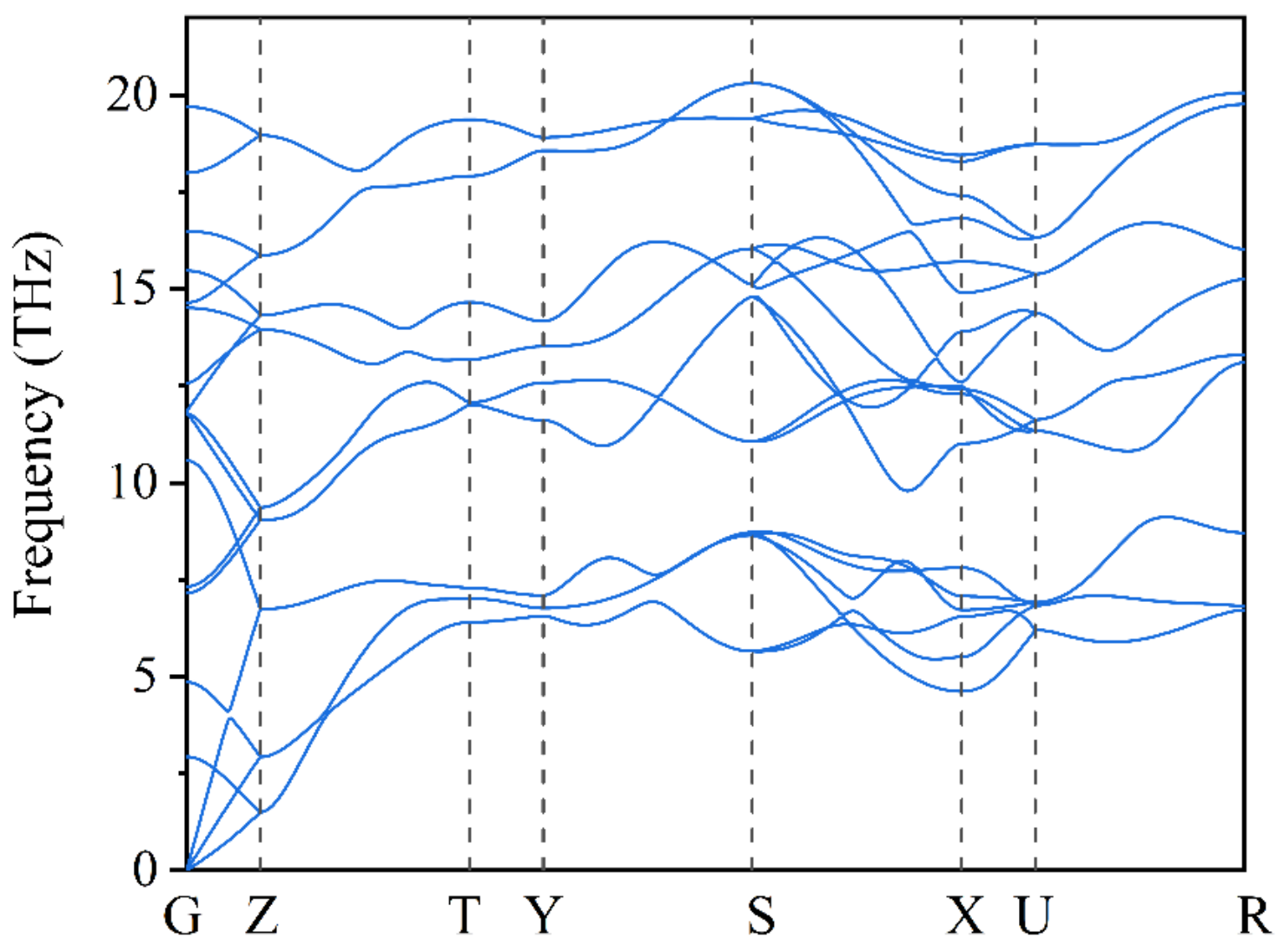

**Figure S9.** Phonon dispersion curves of $Ca_2I$ in *Pmmn* structure at 300 GPa.

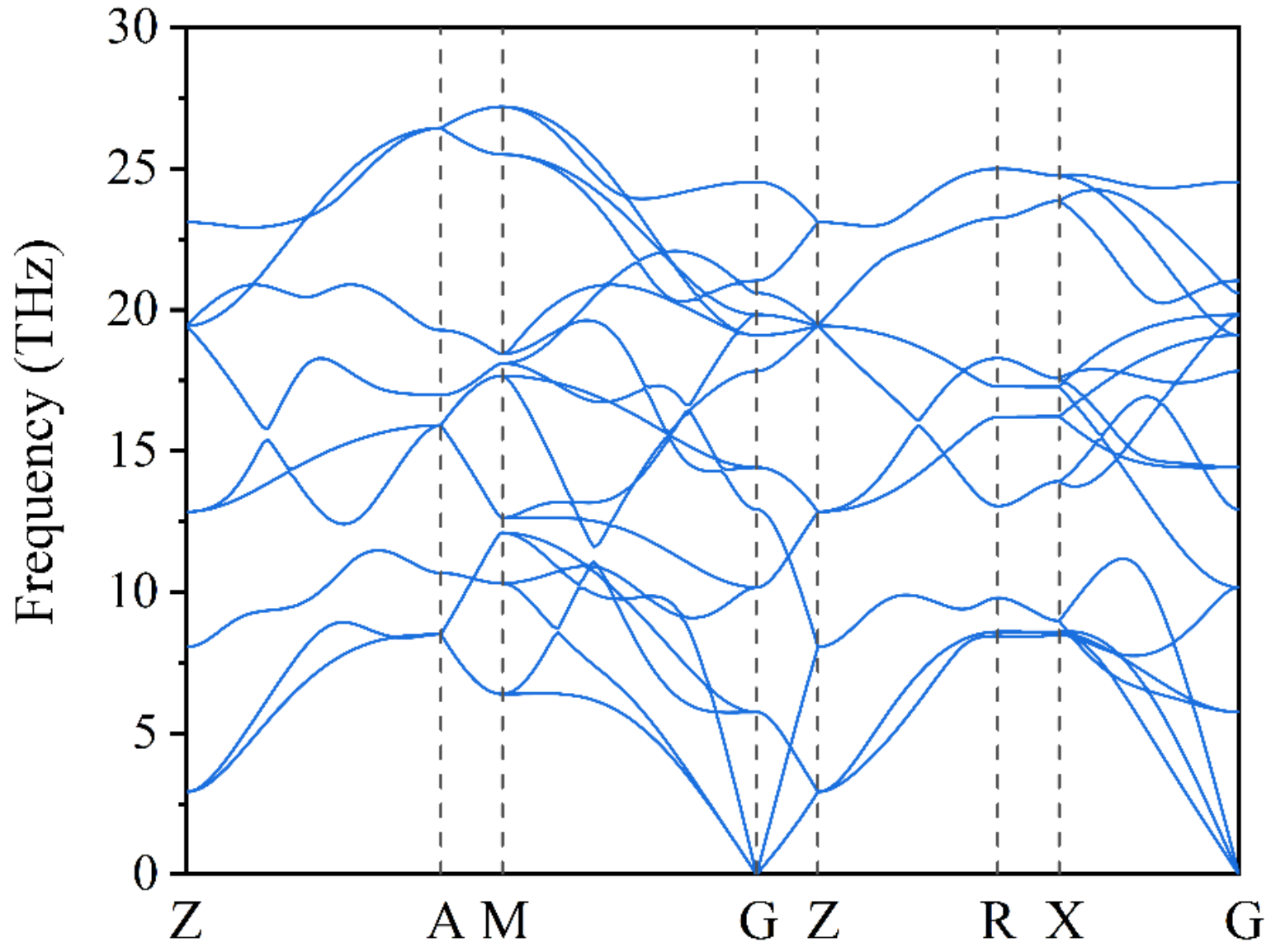

**Figure S10.** Phonon dispersion curves of $Ca_2I$ in *I*4/*mmm* structure at 600 GPa.

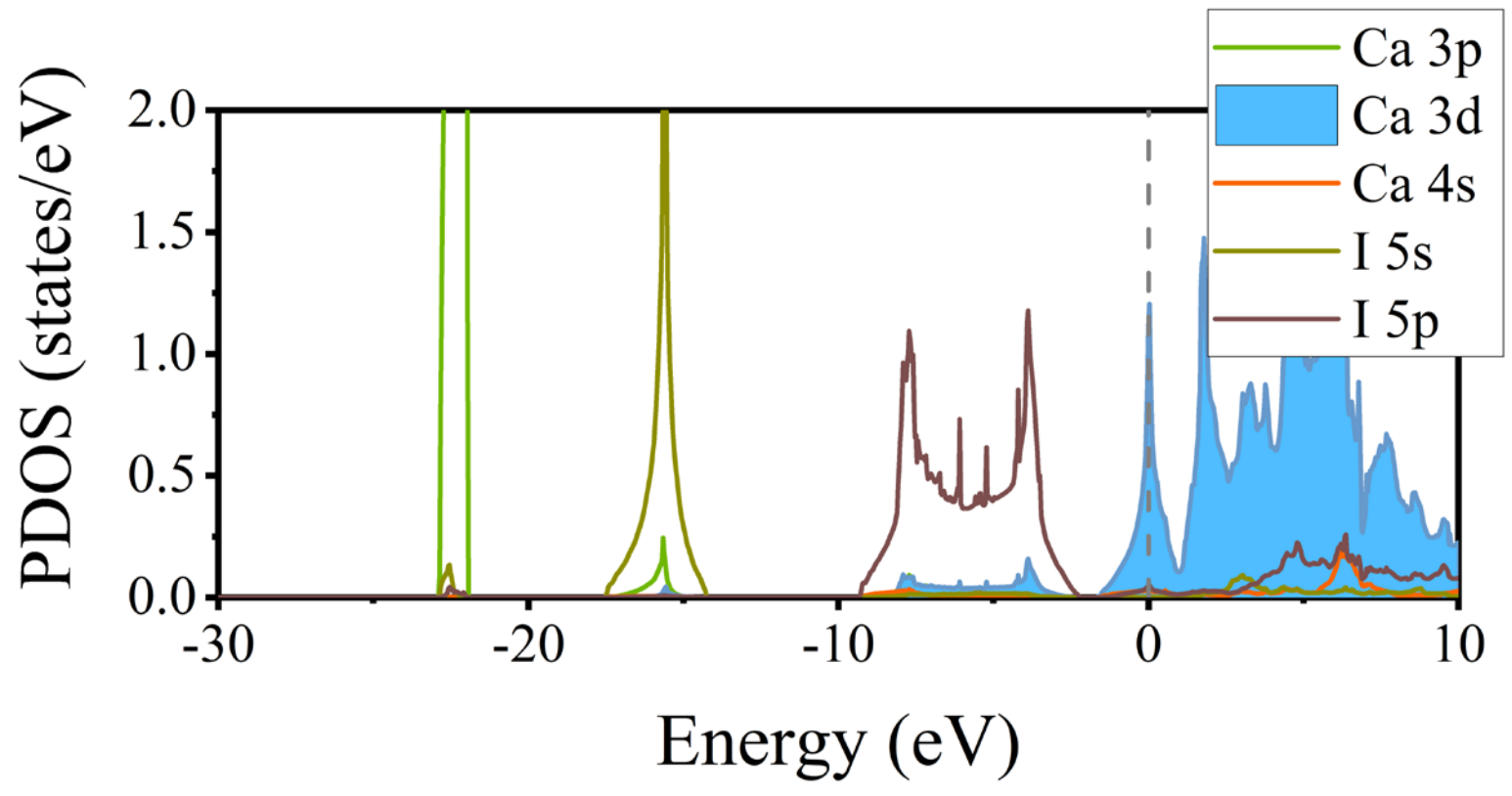

**Figure S11.** Calculated difference of projected density of states (PDOS) of CaI in *Pm*-3*m* structure at 50 GPa.

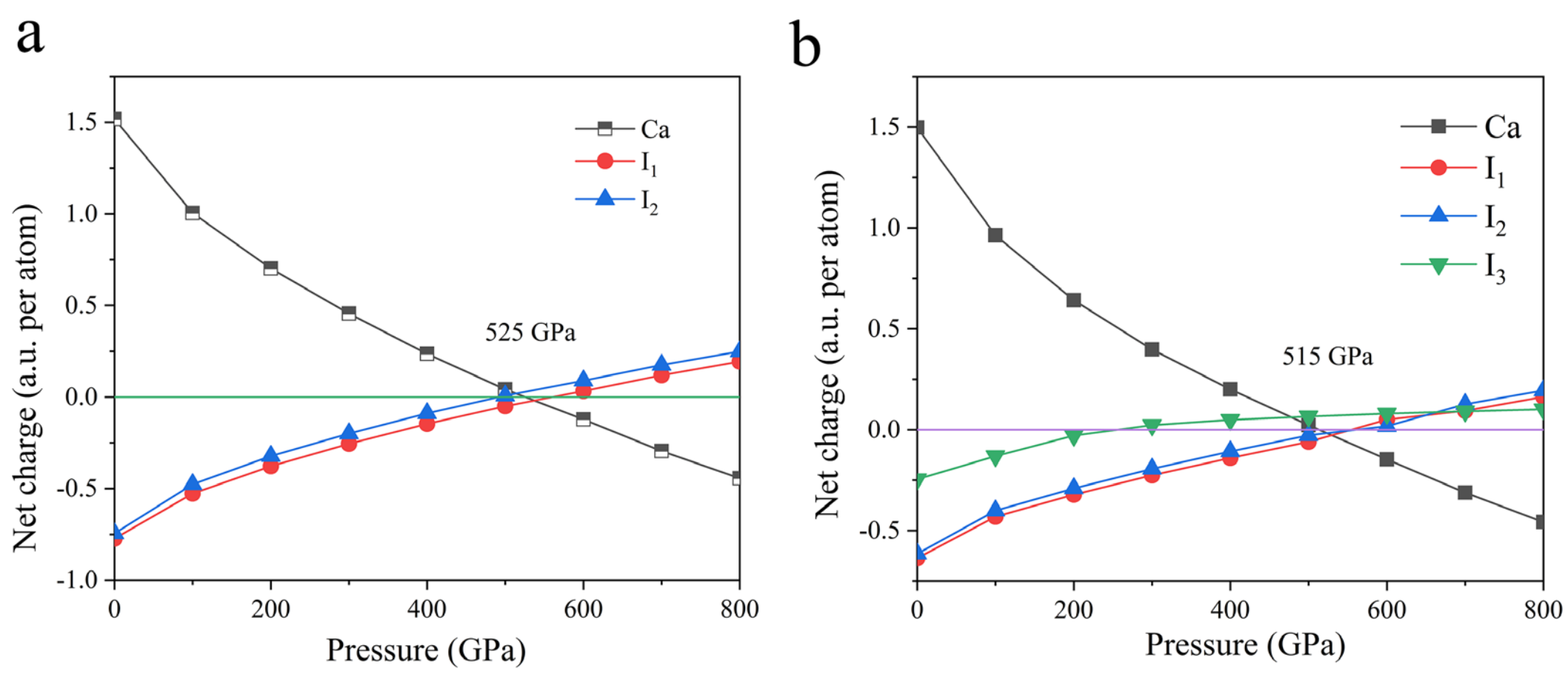

**Figure S12.** (a) The net charge of Ca and I atoms in $CaI_2$ is based on Bader analysis. (b) The net charge of Ca and I atoms in $CaI_3$ is based on Bader analysis.

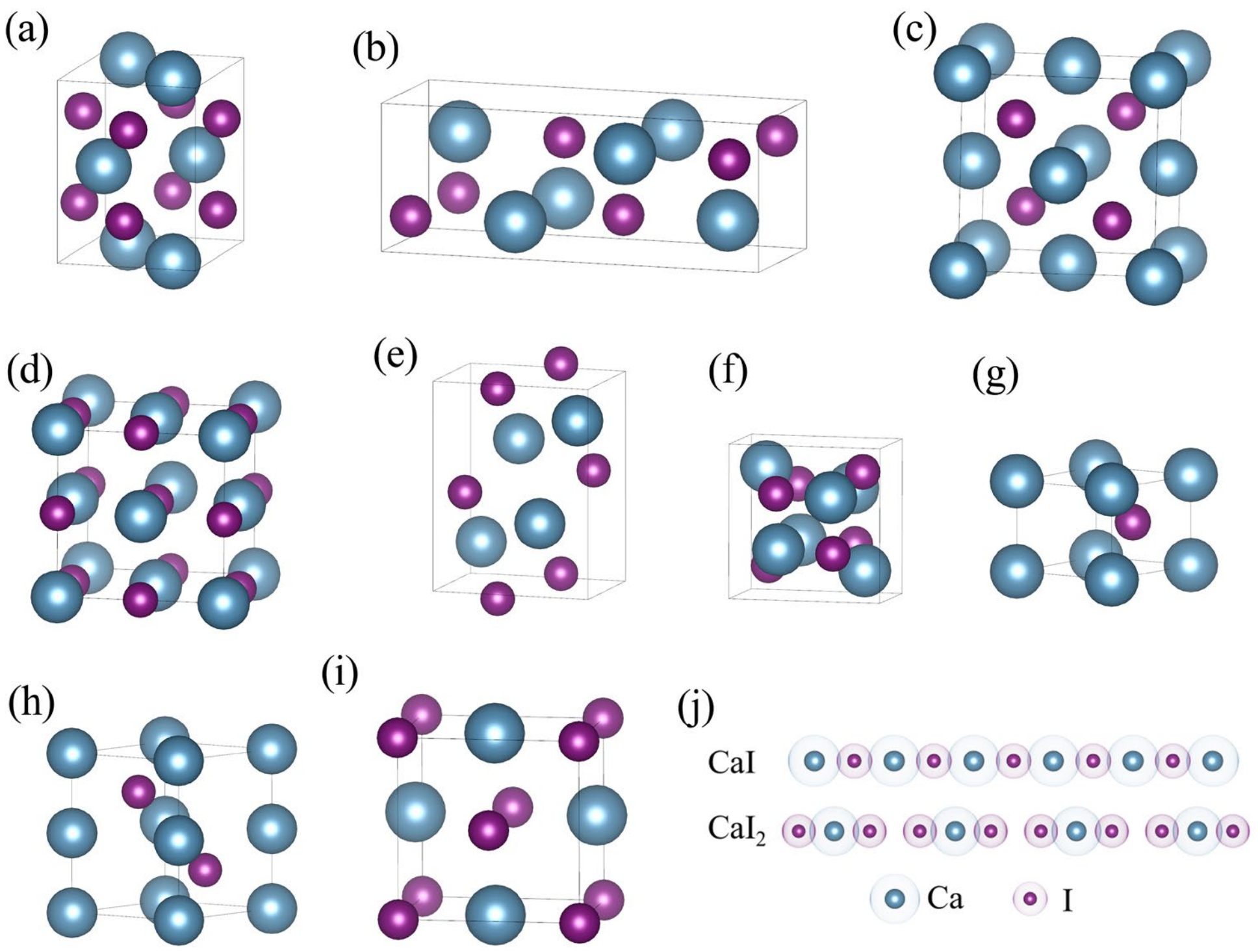

**Figure S13.** Models of various AB-type compounds in Ca-I systems. (a) *C*2/*c*. (b) *Cmcm*. (c) *F*-43*m*. (d) *Fm*-3*m*. (e) *Pnma*. (f) *P*4$_2$/*mnm*. (g) *P*6$_3$/*mc*. (h) *P*-6*m*2. (i) *P*4/*nmm*. (j) 1D model.

## References

(1) Peng, F.; Wang, Y.; Wang, H.; Zhang, Y.; Ma, Y. Stable Xenon Nitride at High Pressures. *Phys. Rev. B* **2015**, *92* (9), 094104. https://doi.org/10.1103/PhysRevB.92.094104.

(2) Li, J.; Geng, Y.; Xu, Z.; Zhang, P.; Garbarino, G.; Miao, M.; Hu, Q.; Wang, X. Mechanochemistry and the Evolution of Ionic Bonds in Dense Silver Iodide. *Journal of the American Chemical Society* **2022**. https://doi.org/10.1021/jacsau.2c00550.

(3) Liu, Y.; Li, J.; Geng, Y.; Xu, Z.; Lv, Y.; Zhang, Z.; Yuan, J.; Wang, X. Pressure-Induced Phase Transitions and Decompositions of Sr-S Compounds. *Physica B: Condensed Matter* **2024**, *681*, 415846.

(4) Xu, Z.; Li, J.; Geng, Y.; Zhang, Z.; Lv, Y.; Zhang, C.; Wang, Q.; Wang, X. Regulation of Ionic Bond in Group IIB Transition Metal Iodides. *Chinese Physics Letters* **2023**, *40* (7), 076201. https://doi.org/10.1088/0256-307X/40/7/076201.

(5) Xu, Z.; Rui, Q.; Geng, Y.; Li, J.; Wang, Q.; Wang, X. Pressure-Induced Decomposition of Cadmium Iodide. *EPL* **2022**, *140* (1), 16003. https://doi.org/10.1209/0295-5075/ac94f4.

(6) Profeta, M.; Benoit, M.; Mauri, F.; Pickard, C. J. First-Principles Calculation of the $^{17}$O NMR Parameters in Ca Oxide and Ca Aluminosilicates: The Partially Covalent Nature of the Ca−O Bond, a Challenge for Density Functional Theory. *J. Am. Chem. Soc.* **2004**, *126* (39), 12628–12635. https://doi.org/10.1021/ja0490830.

(7) Posternak, M.; Baldereschi, A.; Krakauer, H.; Resta, R. Non-Nominal Value of the Dynamical Effective Charge in Alkaline-Earth Oxides. *Phys. Rev. B* **1997**, *55* (24), R15983–R15986. https://doi.org/10.1103/PhysRevB.55.R15983.

(8) Sun, H.; Deng, K.; Kan, E.; Du, Y. Second-Order Jahn–Teller Effect Induced High-Temperature Ferroelectricity in Two-Dimensional $NbO_2$ X (X = I, Br). *Nanoscale Adv.* **2023**, *5* (11), 2979–2985. https://doi.org/10.1039/D3NA00245D.

(9) Yao, Q.-F.; Cai, J.; Tong, W.-Y.; Gong, S.-J.; Wang, J.-Q.; Wan, X.; Duan, C.-G.; Chu, J. H. Manipulation of the Large Rashba Spin Splitting in Polar Two-Dimensional Transition-Metal Dichalcogenides. *Phys. Rev. B* **2017**, *95* (16), 165401. https://doi.org/10.1103/PhysRevB.95.165401.